\documentclass[review]{elsarticle}

\usepackage{hyperref}
\usepackage{siunitx}
\usepackage{booktabs}
\usepackage{xcolor}
\usepackage{float}
\usepackage[normal,raggedright,scriptsize]{subfigure}
\usepackage[normalem]{ulem}

\begin{document}

\begin{frontmatter}

\title{Calculation of force and torque between two arbitrarily oriented circular filaments using  Kalantarov-Zeitlin's method }


\author[innopolis,imt]{Kirill V. Poletkin}
\ead{k.poletkin@innopolis.ru, kirill.poletkin@kit.edu}

\address[innopolis]{Innopolis University, 1, Universitetskaya Str., Innopolis, 420500, Russia}
\address[imt]{The Institute of Microstructure Technology, Karlsruhe Institute of
Technology, Hermann-von-Helmholtz-Platz 1, 76344 Eggenstein-Leopoldshafen, Germany}


\begin{abstract}
In this article, formulas for calculation of force and torque between two circular filaments arbitrarily oriented in space were derived by using Kalantarov-Zeitlin's method. Formulas are presented in an analytical form through  integral expressions, whose kernel function is expressed in terms of the elliptic integrals of the first and second kinds, and provide  an alternative formulation to  Babi\v{c}'s expressions. 
The derived new formulas were validated via comparison  with a series of reference examples. 
 Also,  we obtained additional expressions for calculation of force and torque between two circular filaments by   means of differentiation of Grover’s formula for the mutual inductance between two circular filaments with respect to appropriate  coordinates. These additional expressions allow  to verifying   comprehensively and independently derived new formulas.
\end{abstract}


\end{frontmatter}


\section{Introduction}

Analytical and semi-analytical methods in the calculation of parameters of electrical circuits and force interaction between their elements
play an important role in power transfer, wireless communication, and sensing and actuation, and are applied in different fields of science, including electrical and electronic engineering, medicine, physics, nuclear magnetic resonance, mechatronics and robotics, to name the most prominent. 
A number of efficient  numerical methods implemented in the commercially available software currently provides an accurate and fast solution for the calculation of parameters of electrical circuits. However, analytical methods allow to obtain the result in the form of a final formula with a finite number of input parameters, which when applicable may significantly reduce computation effort. It will also facilitate mathematical analysis, for example, when derivatives of the mutual inductance 
with respect to one or more parameters are required to evaluate electromagnetic forces via the stored magnetic energy, or when optimization is performed.

Analytical methods applied to the calculation of force and torque between two circular filaments, which is closely related to calculation of  the mutual inductance in  such the system, 
is a prime example.  
These methods have been successfully used in an increasing number of applications, including electromagnetic levitation \cite{OkressWroughtonComenetzEtAl1952}, superconducting levitation {\cite{Urman1997,Urman1997a,Coffey2001}}, {magnetic force interaction \cite{Urman2014}}, wireless power transfer \cite{JowGhovanloo2007,SuLiuHui2009,ChuAvestruz2017}, electromagnetic actuation \cite{ShiriShoulaie2009,RavaudLemarquandLemarquand2009,Obata2013,ShalatiPoletkinKorvinkEtAl2018}, micro-machined contactless inductive suspensions \cite{Poletkin2013,Poletkin2014a,Lu2014,PoletkinLuWallrabeEtAl2017b,Vlnieska2020} and hybrid suspensions \cite{Poletkin2012,PoletkinShalatiKorvinkEtAl2018,PoletkinKorvink2018,Poletkin2020}, biomedical applications \cite{TheodoulidisDitchburn2007,SawanHashemiSehilEtAl2009}, topology optimization \cite{KuznetsovGuest2017}, nuclear magnetic resonance \cite{D.I.B.2002,SpenglerWhileMeissnerEtAl2017}, indoor positioning systems \cite{AngelisPaskuAngelisEtAl2015}, navigation sensors \cite{WuJeonMoonEtAl2016}, wireless power transfer systems \cite{Zhang2021} and magneto-inductive wireless communications \cite{Gulbahar2017}.

The calculation of force and torque between two circular filaments with running electrical current is reduced to finding first derivatives of a function of mutual inductance corresponding to  this filament system. 
In 1954,  using the mutual inductance between two coaxial circular filaments  derived by Maxwell \cite[page 340, Art. 701]{Maxwell1954}, C. Snow obtained the formula for calculation attractive force between them   in work \cite{Snow1954}. Also, in the same work \cite{Snow1954} C. Snow presented formulas for calculation of  torque between circular filaments covering the following cases, namely, when axes of circular filaments are intersected, and two concentric circles,  and for calculation of force between  two parallel circles.     The formulas were expressed via series over   the Legendre polynomials. In 1996, Kim et al. developed the expression for calculation of the restoring force in a system of two non-coaxial coils  based on magnetic potential method \cite{Kim1996}. Employing Grover's formula for calculation of mutual inductance between two filament coils \cite{Grover1944},  Babi\v{c} et al. developed the formulas for calculation of force and torque in such filament coil systems, in which circles have the lateral misalignment \cite{Babic2012}  and whose ases are inclined at the same plane \cite{Babic2011,Babic2018}, respectively. The developed formulas were applied to the calculation of force interaction between superconducting magnets and coils having  a rectangular cross-section. In work \cite{Babic2012a}, Babi\v{c} et al. presented new general formulas for calculating the magnetic force between inclined circular filaments placed in any desired position based on two  approaches, namely, Biot-Savart's law and the formula for calculation of  the mutual inductance \cite{BabicSiroisAkyelEtAl2010}.

In this article, new formulas for calculation of force and torque between two circular filaments arbitrarily oriented in space presented in an integral analytical form,  whose kernel function is expressed in terms of the elliptic integrals of the
first and second kinds,  were derived based on Kalantarov-Zeitlin's method and provide  an alternative formulation to  Babi\v{c}'s expressions. Kalantarov and Zeitlin showed that the calculation of mutual inductance between a circular primary filament and any other secondary filament having an arbitrary shape and any desired position with respect to the primary filament can be reduced to a line integral \cite[Sec. 1-12, page 49]{Kalantarov1986}.
Adapting  this method to the case of two circular filaments, the author derived analytical formulas for calculating the mutual inductance between two circular filaments having any desired position with respect to each other in work \cite{Poletkin2019}. Taking first derivatives with respect to the appropriate coordinates of the formulas for calculation of the mutual inductance obtained by means of  Kalantarov-Zeitlin's method, the derivation of these presented formulas for calculation of force and torque was carried out.

Developed new formulas  were successfully verified by the examples taken from  Babi\v{c} et al. work \cite{Babic2012a} and comparison with results of calculation of force and torque performed by expressions derived from   Grover's  formula for calculation of mutual inductance  \cite[page 207, Eq. (179)]{Grover2004}. For the convenience of a reader, all derived formulas including expressions derived by  Grover's  method were programmed by using  the \textit{Matlab} language. The \textit{Matlab} files with the implemented formulas  are available from the author  as   supplementary materials to this article.

\begin{figure}[!t]
  \centering
  \includegraphics[width=2.7in]{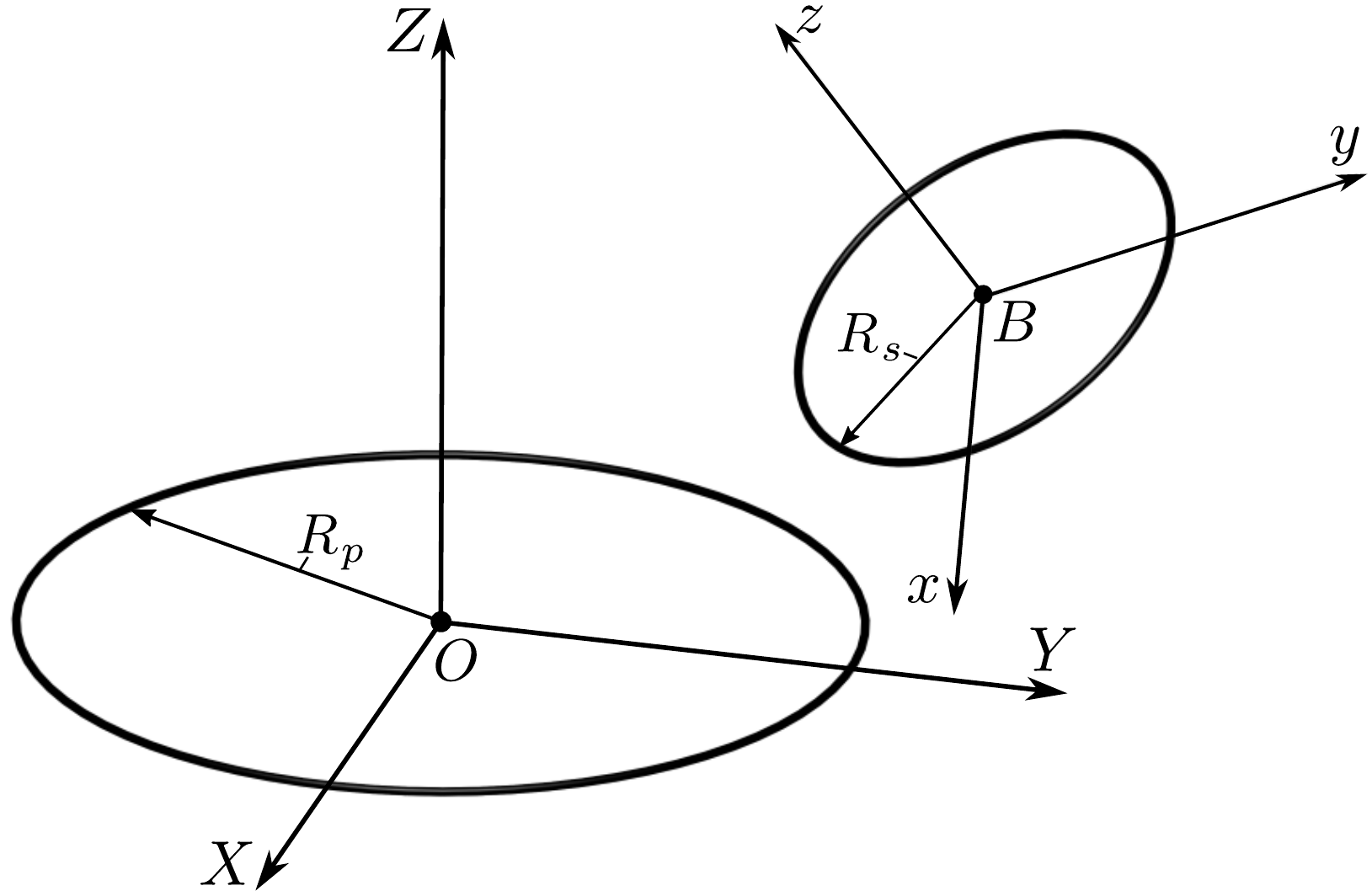}
  \caption{General scheme of arbitrarily positioning two circular filaments with respect to each other.   }\label{fig:scheme}
\end{figure}


\section{Preliminary discussion}
\label{sec:Pleliminary}

Similar to our previous work \cite{Poletkin2019}, the general scheme of arbitrarily positioning of two circular filaments with respect to each other is considered as shown in Fig. \ref{fig:scheme}. 
The primary circular filament (the primary circle) and the secondary circular filament (the secondary circle) have radii of $R_p$ and $R_s$, respectively.  
  A coordinate frame (CF) denoted as $XYZ$ is assigned to the primary circle in such a way that the $Z$ axis is coincident with the circle axis and the $XOY$ plane of the CF lies on  circle's plane, where the origin $O$ corresponds to the centre of primary circle. In turn, the $xyz$ CF is assigned to the secondary circle in a similar way so that its origin $B$ is coincident with the centre of the secondary circle.

\begin{figure*}[!t]
    \centering
     \subfigure[]
    {
    \centering
        \includegraphics[width=1.8in]{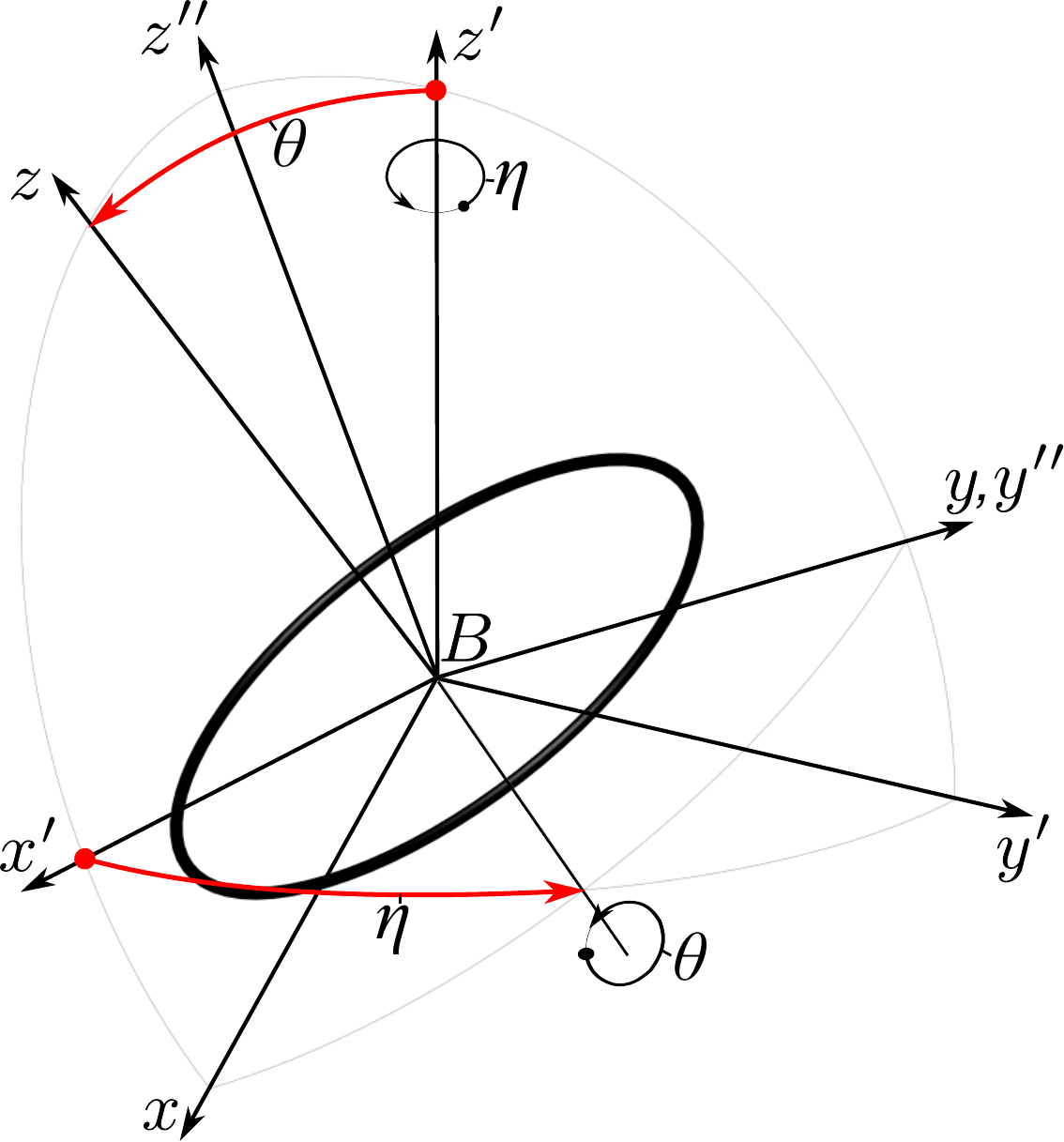}
       \label{fig:Grover angles}
        }\quad
        \subfigure[ ]
    {
    \centering
        \includegraphics[width=1.8in]{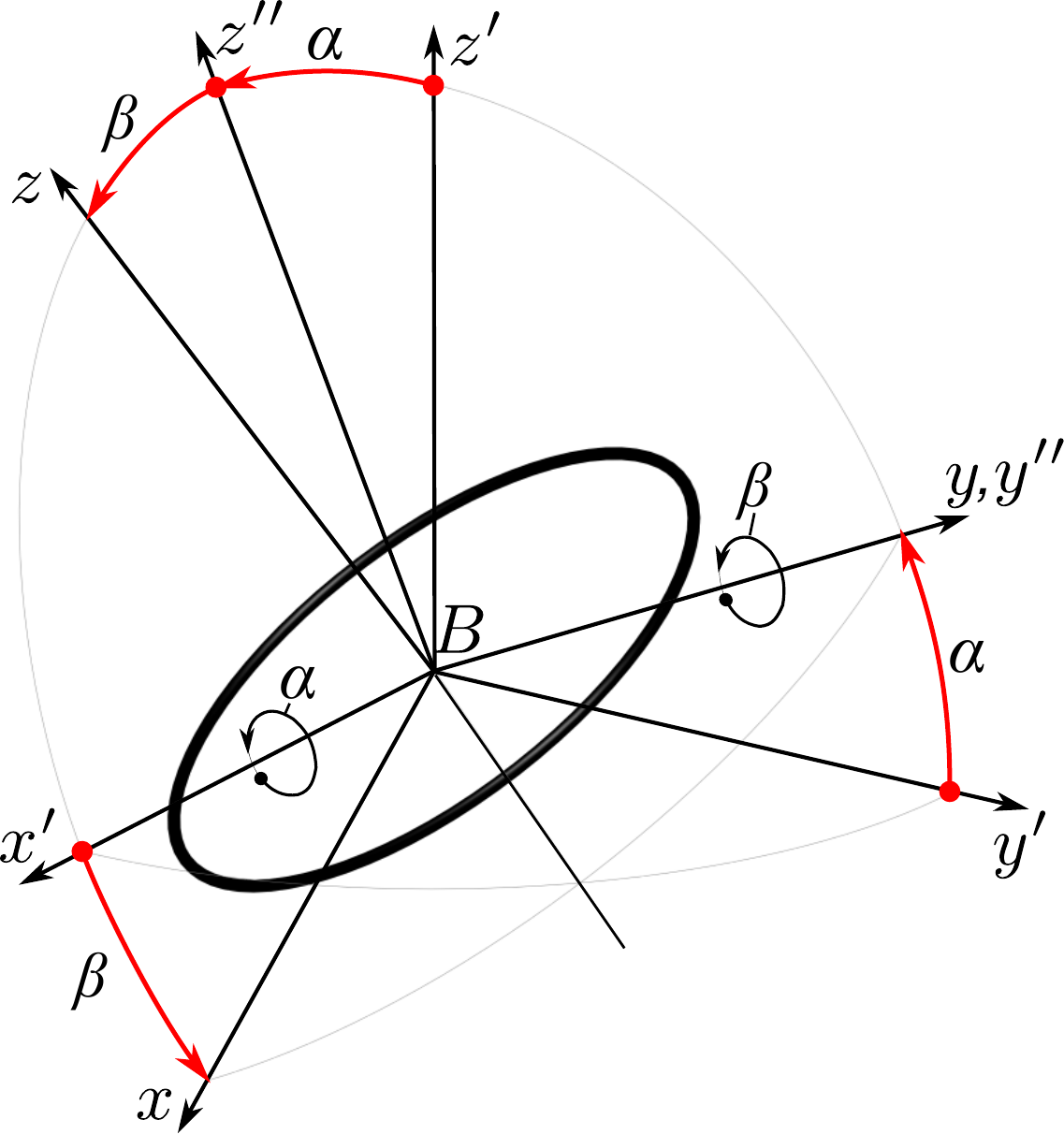}
       \label{fig:Bronyan}
        }\quad
    \caption{Two manners for determining the angular position of the secondary circle with respect to the primary one: $x'y'z'$ is the auxiliary CF the axes of which are parallel to the axes of $XYZ$, respectively; $x''y''z''$ is the auxiliary CF defined in such a way that the $x'$ and $x''$ are coincide, but the $z''$ and $y''$ axis is rotated by the $\alpha$ angle with respect to the $z'$ and $y'$ axis, respectively.  }
    \label{fig:angular position}
\end{figure*}

The linear misalignment of the secondary circle with respect to the primary one is defined by the coordinates of the centre $B$ ($x_B,y_B,z_B$).
The angular misalignment of the secondary circle can be defined by using Grover's  angles \cite[page 207]{Grover2004}. Namely, the angle of $\theta$ and $\eta$ corresponds to the angular rotation around an axis passing through the diameter of the secondary circle, and then the rotation of this axis lying on the surface $x'By'$ around the vertical $z'$ axis, respectively, as it is shown in Figure \ref{fig:angular position}(a).
As an alternative manner to  Grover's angles, the same angular misalignment can be determined through the $\alpha$ and $\beta$ angle, which corresponds to the angular rotation around the $x'$ axis and then around the $y''$ axis, respectively, as it is shown in Figure \ref{fig:angular   position}(b). This additional second  manner  is more convenient in a case of study dynamics and stability issues, for instance, applying to axially symmetric inductive levitation systems \cite{Poletkin2014a,PoletkinLuWallrabeEtAl2017b} in compared with    Grover's manner.  These two pairs of angles have the following relationship with respect to each other such as \cite{Poletkin2019}:
\begin{equation}\label{eq:angles}
  \left\{\begin{array}{l}
   \sin\beta=\sin\eta\sin\theta;\\
   \cos\beta\sin\alpha=\cos\eta\sin\theta.
  \end{array}\right.
\end{equation}

The mutual inductance between these two filaments can be calculated by
the following formulas, which were derived by using Kalantarov-Zeitlin's approach in work \cite{Poletkin2019} for two cases.
Introducing the following dimensionless coordinates:
  \begin{equation}\label{eq:dimensionless_par}
    {\displaystyle {x}=\frac{x_B}{R_s};\; {y}=\frac{y_B}{R_s};\; {z}=\frac{z_B}{R_s}; {s}=\sqrt{{x}^2+{y}^2},} \\
  \end{equation}
for the first case when the $\theta$ angle is lying in interval of $0\leq\theta<\pi/2$, the formula can be written as
\begin{equation}\label{eq:NEW FORMULA}
  M=\frac{\mu_0\sqrt{R_pR_s}}{\pi}\int_{0}^{2\pi}{r}\cdot U\cdot\Phi(k)d\varphi,
\end{equation}
where
\begin{equation}\label{eq:r}
   {r}={r}(\theta,\eta)=\frac{\cos\theta}{\sqrt{\sin^2(\varphi-\eta)+\cos^2\theta\cos^2(\varphi-\eta)}},
\end{equation}
\begin{equation}\label{eq:U}
  U=U({x},{y},\theta,\eta)=\frac{R}{{\rho}^{1.5}}=\frac{{r}+t_1\cdot\cos\varphi+t_2\cdot\sin\varphi}{{\rho}^{1.5}},
\end{equation}
\begin{equation}\label{eq:t and rho}
  \begin{array}{l}
    t_1=t_1({x},{y},\theta,\eta)={x}+0.5\bar{r}^2\tan^2\theta\sin(2(\varphi-\eta))\cdot{y}, \\
     t_2=t_2({x},{y},\theta,\eta)={y}-0.5\bar{r}^2\tan^2\theta\sin(2(\varphi-\eta))\cdot{x},\\
    {\rho}={\rho}({x},{y},\theta,\eta) =\sqrt{{r}^2+2{r}\cdot \left({x}\cos(\varphi)+{y}\sin(\varphi)\right)+{s}^2},
  \end{array}
\end{equation}
\begin{equation}\label{eq:Phi}
   \Phi(k)=\frac{1}{k}\left[\left(1-\frac{k^2}{2}\right)K(k)-E(k)\right],
\end{equation}
and $K(k)$ and $E(k)$ are the complete elliptic functions of the first and second kind, respectively,
and
\begin{equation}\label{eq:k}
\begin{array}{l}
   {\displaystyle k^2=k^2({x},{y},{z},\theta,\eta)=\frac{4\nu{\rho}}{(\nu{\rho}+1)^2+\nu^2{z}_{\lambda}^2},
}\\
 {\displaystyle \nu=R_s/R_p,\;  {z}_{\lambda}={z}+{r}\tan\theta\sin(\varphi-\eta)}.
\end{array}
\end{equation}
For  the second case when the $\theta$ angle is equal to $\pi/2$ and the two filament circles are mutually perpendicular to each other,  the formula becomes
\begin{equation}\label{eq:Singular case}
 \begin{array}{l}
  {\displaystyle M=\frac{\mu_0\sqrt{R_pR_s}}{\pi}\left\{\int_{-1}^{1}U\cdot\Phi(k)d\bar{\ell}\right.}\\
  {\displaystyle  \left.+\int_{1}^{-1}U\cdot\Phi(k)d\bar{\ell}\right\}},
  \end{array}
\end{equation}
where
\begin{equation}\label{eq:U sing c}
  U=U({x},{y},\eta)=\frac{R}{{\rho}^{1.5}}=\frac{t_1-t_2}{{\rho}^{1.5}},
\end{equation}
\begin{equation}\label{eq:t for singular case}
  \begin{array}{l}
    t_1=t_1({x},\eta)=\sin\eta\cdot({x}+\bar{\ell}\cos\eta), \\
     t_2=t_2({y},\eta)=\cos\eta\cdot({y}+\bar{\ell}\sin\eta),\\
     {\rho}={\rho}({x},{y},\eta) =\sqrt{{s}^2+2\bar{\ell}\cdot  \left({x}\cos(\eta)+{y}\sin(\eta)\right)+\bar{\ell}^2},
  \end{array}
\end{equation}
and $\bar{\ell}=\ell/R_s$ is the dimensionless variable. The functions $\Phi(k)$ and $k=k({x},{y},{z},\eta)$ in formula (\ref{eq:Singular case}) have the same structures as defined by Eq. (\ref{eq:Phi}) and (\ref{eq:k}), respectively. Besides that, in the elliptic module $k=k({x},{y},{z},\eta)$, the ${z}_{\lambda}$ function is governed as follows
\begin{equation}\label{eq: Z singular case}
  {z}_{\lambda}={z}\pm\sqrt{1-\bar{\ell}^2},
\end{equation}
 Note that integrating formula (\ref{eq:Singular case}) between $-1$ and $1$, Eq. (\ref{eq: Z singular case}) is calculated with the positive sign and for the other direction the negative sign is taken.

Assuming that the primary and secondary circular filaments carry the currents of $I_p$ and $I_s$, respectively, hence, the magnetic force and torque between these two circular filaments can be calculated by taking the first derivatives of the function of the magnetic energy stored in such the system with respect to the appropriate coordinates. Hence, force can be  calculated by
\begin{equation}\label{eq: general force}
  {\displaystyle  F_g=I_pI_s\frac{\partial M}{\partial g},}
\end{equation}
where $g=x_B$, $y_B$, or $z_B$.
For torque, we can write:
\begin{equation}\label{eq: general torque}
  {\displaystyle  T_g=I_pI_s\frac{\partial M}{\partial g},}
\end{equation}
where $g=\theta$ or $\eta$. Thus, to derive formulas for calculation of force and torque between two arbitrarily oriented circular filaments, the first derivatives of formulas of mutual inductance, namely, represented by Eq. (\ref{eq:NEW FORMULA}) and (\ref{eq:Singular case})   must be taken.



\section{Derivation of Formulas}

In this section the first derivatives of mutual inductance with respect to appropriate coordinates are taken for the two considered above cases separately.

\subsection{The first case: $0\leq\theta<\pi/2$}
\label{sec:second case}
For this case, formula (\ref{eq:NEW FORMULA}) for calculation of mutual inductance is considered. Its kernel is defined as
\begin{equation}\label{eq:kernel NF}
  \mathrm{Kr}={r}\cdot U\cdot\Phi(k).
\end{equation}
Finding the first derivatives of mutual inductance (\ref{eq:NEW FORMULA}) is reduced to taking the derivatives of kernel function. 
Following this, firstly five derivatives of the kernel with respect to the defined five coordinates are obtained as follows.

According to the definitions of functions $r$, $U$, $\Phi(k)$ and $k$ given in Eq. (\ref{eq:r}), (\ref{eq:U}), (\ref{eq:Phi}) and (\ref{eq:k}), respectively, the $x_B$-derivative of kernel $ \mathrm{Kr}$ can be written as
\begin{equation}\label{eq:first x-der of kernel NF}
  {\displaystyle  \frac{\partial\mathrm{Kr}}{\partial x_B}=\frac{\partial\mathrm{Kr}}{\partial x}\frac{1}{R_s}=\frac{r}{R_s}\cdot\left[\frac{\partial U}{\partial x} \cdot\Phi(k)+U\cdot\frac{d \Phi(k)}{d k} \cdot\frac{\partial k}{\partial x}\right],}
\end{equation}
where
\begin{equation}\label{eq:dU-x}
 \begin{array}{l}
  {\displaystyle  \frac{\partial U}{\partial x}=\left({\displaystyle\frac{\partial R}{\partial x}}\cdot\rho-1.5\cdot R\cdot\frac{\partial \rho}{\partial x}\right)\bigg/{{\rho}^{2.5}},}\\
  {\displaystyle \frac{\partial R}{\partial x}=\frac{\partial t_1}{\partial x}\cdot\cos\varphi+\frac{\partial t_2}{\partial x}\cdot\sin\varphi},\\
   {\displaystyle \frac{\partial t_1}{\partial x}=1,\;\frac{\partial t_2}{\partial x}=-0.5\bar{r}^2\tan^2\theta\sin(2(\varphi-\eta))},\\
    {\displaystyle \frac{\partial \rho}{\partial x}=\left(r\cdot\cos\varphi +x\right)\big/{{\rho}},}
   \end{array}
\end{equation}
\begin{equation}\label{eq:dPhi}
 \frac{d \Phi(k)}{d k}=\frac{1}{k^2}\left[\frac{2-k^2}{2(1-k^2)}E(k)-K(k)\right],
\end{equation}
\begin{equation}\label{eq:dk-x}
   {\displaystyle \frac{\partial k}{\partial x}=\frac{2/k-k(\nu{\rho}+1)}{(\nu{\rho}+1)^2+\nu^2{z}_{\lambda}^2}\cdot\nu\frac{\partial \rho}{\partial x}.
}
\end{equation}
For the $y_B$-derivative of kernel $ \mathrm{Kr}$, we have
\begin{equation}\label{eq:first y-der of kernel NF}
  {\displaystyle  \frac{\partial\mathrm{Kr}}{\partial y_B}=\frac{\partial\mathrm{Kr}}{\partial y}\frac{1}{R_s}=\frac{r}{R_s}\cdot\left[\frac{\partial U}{\partial y} \cdot\Phi(k)+U\cdot\frac{d \Phi(k)}{d k} \cdot\frac{\partial k}{\partial y}\right],}
\end{equation}
where
\begin{equation}\label{eq:dU-y}
 \begin{array}{l}
  {\displaystyle  \frac{\partial U}{\partial y}=\left({\displaystyle\frac{\partial R}{\partial y}}\cdot\rho-1.5\cdot R\cdot\frac{\partial \rho}{\partial y}\right)\bigg/{{\rho}^{2.5}},}\\
  {\displaystyle \frac{\partial R}{\partial y}=\frac{\partial t_1}{\partial y}\cdot\cos\varphi+\frac{\partial t_2}{\partial y}\cdot\sin\varphi},\\
   {\displaystyle \frac{\partial t_1}{\partial y}= 0.5\bar{r}^2\tan^2\theta\sin(2(\varphi-\eta)),\;\frac{\partial t_2}{\partial y}=1},\\
    {\displaystyle \frac{\partial \rho}{\partial y}=\left(r\cdot\sin\varphi +y\right)\big/{{\rho}}.}
   \end{array}
\end{equation}
The derivative of elliptic module $k$ with respect to $y$ has a similar form to Eq. (\ref{eq:dk-x}), but in the later one  the partial derivative, $  {\displaystyle \frac{\partial \rho}{\partial x}}$, must be replaced by  ${\displaystyle \frac{\partial \rho}{\partial y}}$ defined in (\ref{eq:dU-y}).
The $z_B$-derivative of kernel $ \mathrm{Kr}$ is
\begin{equation}\label{eq:first z-der of kernel NF}
  {\displaystyle  \frac{\partial\mathrm{Kr}}{\partial z_B}=\frac{\partial\mathrm{Kr}}{\partial z}\frac{1}{R_s}=\frac{r}{R_s}\cdot U\cdot\frac{d \Phi(k)}{d k} \cdot\frac{\partial k}{\partial z},}
\end{equation}
where
\begin{equation}\label{eq:dk-z}
   {\displaystyle \frac{\partial k}{\partial z}=-\sqrt{4\nu\rho}\cdot\frac{\nu^2{z}_{\lambda}}{\left((\nu{\rho}+1)^2+\nu^2{z}_{\lambda}^2\right)^{3/2}}.
}
\end{equation}
The $\theta$-derivative of kernel $ \mathrm{Kr}$ becomes as follows
\begin{equation}\label{eq:first theta-der of kernel NF}
  {\displaystyle  \frac{\partial\mathrm{Kr}}{\partial \theta}=\left[\frac{\partial r}{\partial\theta}\cdot U+r\cdot\frac{\partial U}{\partial  \theta}\right]\cdot\Phi(k)+r\cdot U \cdot \frac{d \Phi(k)}{d k} \cdot\frac{\partial k}{\partial \theta},}
\end{equation}
where
\begin{equation}\label{eq:dr-theta}
   \frac{\partial r}{\partial \theta}=-\frac{{\sin\left(\eta -\varphi \right)}^2\,\sin\left(\theta \right)}{{\left({\sin\left(\eta -\varphi \right)}^2+{\cos\left(\theta \right)}^2{\cos\left(\eta -\varphi \right)}^2\right)}^{3/2}},
\end{equation}
\begin{equation}\label{eq:dU-theta}
 \begin{array}{l}
  {\displaystyle  \frac{\partial U}{\partial \theta}=\left({\displaystyle\frac{\partial R}{\partial \theta}}\cdot\rho-1.5\cdot R\cdot\frac{\partial \rho}{\partial \theta}\right)\bigg/{{\rho}^{2.5}},}\\
  {\displaystyle \frac{\partial R}{\partial \theta}=\frac{\partial r}{\partial \theta}+\frac{\partial t_1}{\partial  \theta}\cdot\cos\varphi+\frac{\partial t_2}{\partial  \theta}\cdot\sin\varphi},\\
   {\displaystyle \frac{\partial t_1}{\partial \theta}=r\cdot y\cdot\sin\left( 2\,\varphi- 2\,\eta\right)\,\mathrm{tan}\left(\theta \right)\left[\frac{\partial r}{\partial \theta}\cdot\mathrm{tan}\left(\theta \right)+\frac{r}{{\cos\left(\theta \right)}^2}\right]},\\
   {\displaystyle \frac{\partial t_2}{\partial \theta}=-r\cdot x\cdot\sin\left( 2\,\varphi- 2\,\eta\right)\,\mathrm{tan}\left(\theta \right)\left[\frac{\partial r}{\partial \theta}\cdot\mathrm{tan}\left(\theta \right)+\frac{r}{{\cos\left(\theta \right)}^2}\right]},\\
    {\displaystyle \frac{\partial \rho}{\partial \theta}=\frac{r+ y\cdot\sin\varphi +x\cdot\cos\varphi}{\rho}\cdot\frac{\partial r}{\partial \theta}},
   \end{array}
\end{equation}
\begin{equation}\label{eq:dk-theta}
\begin{array}{l}
   {\displaystyle \frac{\partial k}{\partial \theta}= \frac{{\displaystyle\left[2/k-k(\nu{\rho}+1)\right]\cdot\nu\frac{\partial \rho}{\partial \theta}-k\cdot\nu^2{z}_{\lambda}\frac{\partial {z}_{\lambda}}{\partial \theta}}}{(\nu{\rho}+1)^2+\nu^2{z}_{\lambda}^2}.
}\\
 {\displaystyle   \frac{\partial {z}_{\lambda}}{\partial \theta}=\sin(\varphi-\eta)\left[\frac{\partial r}{\partial \theta}\cdot\mathrm{tan}\left(\theta \right)+\frac{r}{{\cos\left(\theta \right)}^2}\right]}.
\end{array}
\end{equation}
The $\eta$-derivative of kernel $ \mathrm{Kr}$ can be written as
\begin{equation}\label{eq:first eta-der of kernel NF}
  {\displaystyle  \frac{\partial\mathrm{Kr}}{\partial \eta}=\left[\frac{\partial r}{\partial\eta}\cdot U+r\cdot\frac{\partial U}{\partial \eta}\right]\cdot\Phi(k)+r\cdot U \cdot \frac{d \Phi(k)}{d k} \cdot\frac{\partial k}{\partial \eta},}
\end{equation}
where
\begin{equation}\label{eq:dr-eta}
   \frac{\partial r}{\partial \eta}=\frac{{\sin\left(\varphi -\eta \right)}\,{\cos\left(\varphi - \eta\right)}\,\cos\left(\theta \right)(1-{\cos\left(\theta \right)}^2)}{{\left({\sin\left(\eta -\varphi \right)}^2+{\cos\left(\theta \right)}^2{\cos\left(\eta -\varphi \right)}^2\right)}^{3/2}},
\end{equation}
\begin{equation}\label{eq:dU-eta}
 \begin{array}{l}
  {\displaystyle  \frac{\partial U}{\partial \eta}=\left({\displaystyle\frac{\partial R}{\partial \eta}}\cdot\rho-1.5\cdot R\cdot\frac{\partial \rho}{\partial \eta}\right)\bigg/{{\rho}^{2.5}},}\\
  {\displaystyle \frac{\partial R}{\partial \eta}=\frac{\partial r}{\partial \eta}+\frac{\partial t_1}{\partial  \eta}\cdot\cos\varphi+\frac{\partial t_2}{\partial  \eta}\cdot\sin\varphi},\\
   {\displaystyle \frac{\partial t_1}{\partial \eta}={\mathrm{tan}\left(\theta \right)}^2\cdot r\cdot y\cdot\left[\frac{\partial r}{\partial \eta}\cdot\sin\left( 2\,\varphi- 2\,\eta\right)-r\cdot\cos\left( 2\,\varphi- 2\,\eta\right)\right]},\\
   {\displaystyle \frac{\partial t_2}{\partial \eta}=-{\mathrm{tan}\left(\theta \right)}^2\cdot r\cdot x\cdot\left[\frac{\partial r}{\partial \eta}\cdot\sin\left( 2\,\varphi- 2\,\eta\right)-r\cdot\cos\left( 2\,\varphi- 2\,\eta\right)\right]},\\
    {\displaystyle \frac{\partial \rho}{\partial \eta}=\frac{r+ y\cdot\sin\varphi +x\cdot\cos\varphi}{\rho}\cdot\frac{\partial r}{\partial \eta}},
   \end{array}
\end{equation}
\begin{equation}\label{eq:dk-eta}
\begin{array}{l}
   {\displaystyle \frac{\partial k}{\partial \eta}= \frac{{\displaystyle\left[2/k-k(\nu{\rho}+1)\right]\cdot\nu\frac{\partial \rho}{\partial \eta}-k\cdot\nu^2{z}_{\lambda}\frac{\partial {z}_{\lambda}}{\partial \eta}}}{(\nu{\rho}+1)^2+\nu^2{z}_{\lambda}^2},
}\\
 {\displaystyle   \frac{\partial {z}_{\lambda}}{\partial \eta}={\mathrm{tan}\left(\theta \right)}\cdot\left[\frac{\partial r}{\partial \eta}\cdot\sin\left( \varphi- \eta\right)-r\cdot\cos\left( \varphi- \eta\right)\right]}.
\end{array}
\end{equation}

Accounting for (\ref{eq:first x-der of kernel NF}) and (\ref{eq:first y-der of kernel NF}), the formulas for calculation the first derivatives of mutual inductance between two circular filaments   arbitrarily   oriented in space  relative to the $x$ and $y$ axis  can be written as
\begin{equation}\label{eq: x and y first der of MI}
  {\displaystyle  \frac{\partial M}{\partial g}=\frac{\mu_0}{\pi}\sqrt{\frac{R_p}{R_s}}\int_{0}^{2\pi}r\cdot\left[\frac{\partial U}{\partial g} \cdot\Phi(k)+U\cdot\frac{d \Phi(k)}{d k} \cdot\frac{\partial k}{\partial g}\right]d\varphi,}
\end{equation}
where $g=x_B$, and $y_B$, the derivatives of functions $U $ and $k$ with respect to appropriate coordinates are defined in  (\ref{eq:first x-der of kernel NF}), and (\ref{eq:first y-der of kernel NF}) for $x_B$ and $y_B$ coordinate, respectively. The formula for calculation of the first derivative of the mutual inductance (\ref{eq:NEW FORMULA}) with respect to $z_B$ by taking into account Eq. (\ref{eq:first z-der of kernel NF}) becomes
\begin{equation}\label{eq: z first der of MI}
  {\displaystyle  \frac{\partial M}{\partial z_B}=\frac{\mu_0}{\pi}\sqrt{\frac{R_p}{R_s}}\int_{0}^{2\pi}r\cdot U\cdot\frac{d \Phi(k)}{d k} \cdot\frac{\partial k}{\partial z} d\varphi.}
\end{equation}
The first derivatives of the mutual inductance (\ref{eq:NEW FORMULA}) with respect to angular coordinates by accounting for (\ref{eq:first theta-der of kernel NF}) and (\ref{eq:first eta-der of kernel NF}) are
\begin{equation}\label{eq: theta and eta first der of MI}
  {\displaystyle  \frac{\partial M}{\partial g}=\frac{\mu_0\sqrt{{R_p}{R_s}}}{\pi}\int_{0}^{2\pi}\left[\frac{\partial r}{\partial g}\cdot U+r\cdot\frac{\partial U}{\partial g}\right]\cdot\Phi(k)+r\cdot U \cdot \frac{d \Phi(k)}{d k} \cdot\frac{\partial k}{\partial g} d\varphi,}
\end{equation}
where $g=\theta$, and $\eta$. Substituting (\ref{eq: x and y first der of MI}) and (\ref{eq: z first der of MI}) into Eq. (\ref{eq: general force}), the force between two filaments arbitrarily  oriented in space relative to the $X$, $Y$ and $Z$ axis   and carrying electrical current $I_p$ and $I_s$ can be calculated. Substituting (\ref{eq: theta and eta first der of MI}) into Eq. (\ref{eq: general torque}), the torque acting on these two circular filaments can be estimated.

\subsection{The second case: $\theta=\pi/2$}
For the second case, formula (\ref{eq:Singular case}) for calculation of mutual inductance is used. Its kernel is defined as
\begin{equation}\label{eq:kernel SC}
  \mathrm{Kr}=U\cdot\Phi(k).
\end{equation}
Then, the $x_B$ and $y_B$ -derivative of the kernel (\ref{eq:kernel SC}) are
\begin{equation}\label{eq:first x and y-der of kernel SC}
  {\displaystyle  \frac{\partial\mathrm{Kr}}{\partial g}=\frac{\partial\mathrm{Kr}}{\partial g}\frac{1}{R_s}=\frac{1}{R_s}\cdot\left[\frac{\partial U}{\partial g} \cdot\Phi(k)+U\cdot\frac{d \Phi(k)}{d k} \cdot\frac{\partial k}{\partial g}\right],}
\end{equation}
where  $g=x_B$, and $y_B$,
\begin{equation}\label{eq:dU-g SC}
 \begin{array}{l}
  {\displaystyle  \frac{\partial U}{\partial g}=\left({\displaystyle\frac{\partial R}{\partial g}}\cdot\rho-1.5\cdot R\cdot\frac{\partial \rho}{\partial g}\right)\bigg/{{\rho}^{2.5}},}\\
   {\displaystyle \frac{\partial k}{\partial g}=\frac{2/k-k(\nu{\rho}+1)}{(\nu{\rho}+1)^2+\nu^2{z}_{\lambda}^2}\cdot\nu\frac{\partial \rho}{\partial g},
}
   \end{array}
\end{equation}
for the $x$-derivatives of $R$ and $\rho$:
\begin{equation}\label{eq:dR and drho-x SC}
 \begin{array}{l}
   {\displaystyle \frac{\partial R}{\partial x}=\frac{\partial t_1}{\partial x}=\sin\eta},\\
   {\displaystyle \frac{\partial \rho}{\partial x}=\frac{x+\bar{\ell}\cos\eta}{\rho}},
   \end{array}
\end{equation}
for the $y$-derivatives of $R$ and $\rho$:
\begin{equation}\label{eq:dR and drho-y SC}
 \begin{array}{l}
   {\displaystyle \frac{\partial R}{\partial y}=\frac{\partial t_2}{\partial y}=\cos\eta},\\
   {\displaystyle \frac{\partial \rho}{\partial y}=\frac{y+\bar{\ell}\sin\eta}{\rho}}.
   \end{array}
\end{equation}
The $z_B$-derivative is
\begin{equation}\label{eq:first z-der of kernel SC}
  {\displaystyle  \frac{\partial\mathrm{Kr}}{\partial z_B}=\frac{\partial\mathrm{Kr}}{\partial z}\frac{1}{R_s}=\frac{1}{R_s}\cdot U\cdot\frac{d \Phi(k)}{d k} \cdot\frac{\partial k}{\partial z},}
\end{equation}
where the partial derivative of elliptic module $k$ with respect to $z$ has the same structure as in Eq. (\ref{eq:dk-z}). We can write the following equation for the $\eta$-derivative of the kernel:
\begin{equation}\label{eq:first eta-der of kernel SC}
  {\displaystyle  \frac{\partial\mathrm{Kr}}{\partial \eta}=\frac{\partial U}{\partial \eta} \cdot\Phi(k)+U\cdot\frac{d \Phi(k)}{d k} \cdot\frac{\partial k}{\partial \eta},}
\end{equation}
where
\begin{equation}\label{eq:dU-eta SC}
 \begin{array}{l}
  {\displaystyle  \frac{\partial U}{\partial \eta}=\left({\displaystyle\frac{\partial R}{\partial \eta}}\cdot\rho-1.5\cdot R\cdot\frac{\partial \rho}{\partial \eta}\right)\bigg/{{\rho}^{2.5}},}\\
   {\displaystyle \frac{\partial R}{\partial \eta}=\frac{\partial t_1}{\partial  \eta}-\frac{\partial t_2}{\partial  \eta}=x\cdot\cos\eta+y\cdot\sin\eta},\\
   {\displaystyle \frac{\partial \rho}{\partial \eta}=\frac{ y\cdot\cos\eta -x\cdot\sin\eta}{\rho}\cdot \bar{\ell},}\\
   {\displaystyle \frac{\partial k}{\partial \eta}=\frac{2/k-k(\nu{\rho}+1)}{(\nu{\rho}+1)^2+\nu^2{z}_{\lambda}^2}\cdot\nu\frac{\partial \rho}{\partial \eta}.
}
   \end{array}
\end{equation}

Hence, replacing the kernel in formula (\ref{eq:Singular case}) by Eq. (\ref{eq:first x and y-der of kernel SC}),  the first derivatives of the formula  for calculation of the mutual inductance in the case, when the circular filaments are mutually perpendicular to each other with respect to variables of $x_B$ and $y_B$ can be written as
\begin{equation}\label{eq:first der of SC x and y}
 \begin{array}{l}
  {\displaystyle \frac{\partial M}{\partial g}=\frac{\mu_0}{\pi}\sqrt{\frac{R_p}{R_s}}\left\{\int_{-1}^{1}\left[\frac{\partial U}{\partial g} \cdot\Phi(k)+U\cdot\frac{d \Phi(k)}{d k} \cdot\frac{\partial k}{\partial g}\right]d\bar{\ell}\right.}\\
  {\displaystyle+\left.\int_{1}^{-1}\left[\frac{\partial U}{\partial g} \cdot\Phi(k)+U\cdot\frac{d \Phi(k)}{d k} \cdot\frac{\partial k}{\partial g}\right]d\bar{\ell}\right\}},
  \end{array}
\end{equation}
where  $g=x_B$, and $y_B$.   Accounting for (\ref{eq:first z-der of kernel SC}), the first derivative of  formula (\ref{eq:Singular case}) with respect to $z_B$ becomes
\begin{equation}\label{eq:first der of SC z}
 \begin{array}{l}
  {\displaystyle \frac{\partial M}{\partial z_B}=\frac{\mu_0}{\pi}\sqrt{\frac{R_p}{R_s}}\left\{\int_{-1}^{1}U\cdot\frac{d \Phi(k)}{d k} \cdot\frac{\partial k}{\partial z}d\bar{\ell}\right.}\\
  {\displaystyle  +\left.\int_{1}^{-1}U\cdot\frac{d \Phi(k)}{d k} \cdot\frac{\partial k}{\partial z}d\bar{\ell}\right\}}.
  \end{array}
\end{equation}
The $\eta$-derivative of formula (\ref{eq:Singular case}) by taking into account (\ref{eq:first eta-der of kernel SC}) can be written as follows
\begin{equation}\label{eq:first der of SC eta}
  \begin{array}{l}
  {\displaystyle \frac{\partial M}{\partial \eta}=\frac{\mu_0\sqrt{{R_p}{R_s}}}{\pi}\left\{\int_{-1}^{1}\left[\frac{\partial U}{\partial \eta} \cdot\Phi(k)+U\cdot\frac{d \Phi(k)}{d k} \cdot\frac{\partial k}{\partial \eta}\right]d\bar{\ell}\right.}\\
  {\displaystyle+\left.\int_{1}^{-1}\left[\frac{\partial U}{\partial \eta} \cdot\Phi(k)+U\cdot\frac{d \Phi(k)}{d k} \cdot\frac{\partial k}{\partial \eta}\right]d\bar{\ell}\right\}}.
  \end{array}
\end{equation}
Now the force and torque for this particular case of the configuration of the filament system can be calculated by substituting (\ref{eq:first der of SC x and y}), (\ref{eq:first der of SC z}) and (\ref{eq:first der of SC eta}) into (\ref{eq: general force}) and (\ref{eq: general torque}), respectively.

Since the obtained formulas are intuitively understandable for application, they can be easily programmed. 
For this purpose, the \textit{Matlab} language was used. The \textit{Matlab} files with the implemented formulas (\ref{eq: x and y first der of MI}), (\ref{eq: z first der of MI}), (\ref{eq: theta and eta first der of MI}), (\ref{eq:first der of SC x and y}), (\ref{eq:first der of SC z}) and (\ref{eq:first der of SC eta}) are available from the author  as   supplementary materials to this article. Also,  the   developed formulas can be rewritten through the pair of the angle $\alpha$ and $\beta$.

\section{ Examples of Calculation. Numerical Verification }
In this section,  developed new formulas (\ref{eq: x and y first der of MI}), (\ref{eq: z first der of MI}), (\ref{eq: theta and eta first der of MI}), (\ref{eq:first der of SC x and y}), (\ref{eq:first der of SC z}) and (\ref{eq:first der of SC eta}) are verified by the examples taken from  Babi\v{c} et al. work \cite{Babic2012a} and comparison with results of calculation of force and torque performed by expressions derived from   Grover's  formula for calculation of mutual inductance  \cite[page 207, Eq. (179)]{Grover2004}. The derivations of these expressions are shown in \ref{app:Grover}. In all examples bellow, it is assumed that the carrying currents in both coils are equal to one ampere ($I_p=I_s=\SI{1}{\ampere}$).
All  calculations for considered cases proved the robustness and efficiency of developed formulas.

\begin{figure}[!t]
  \centering
  \includegraphics[width=2.8in]{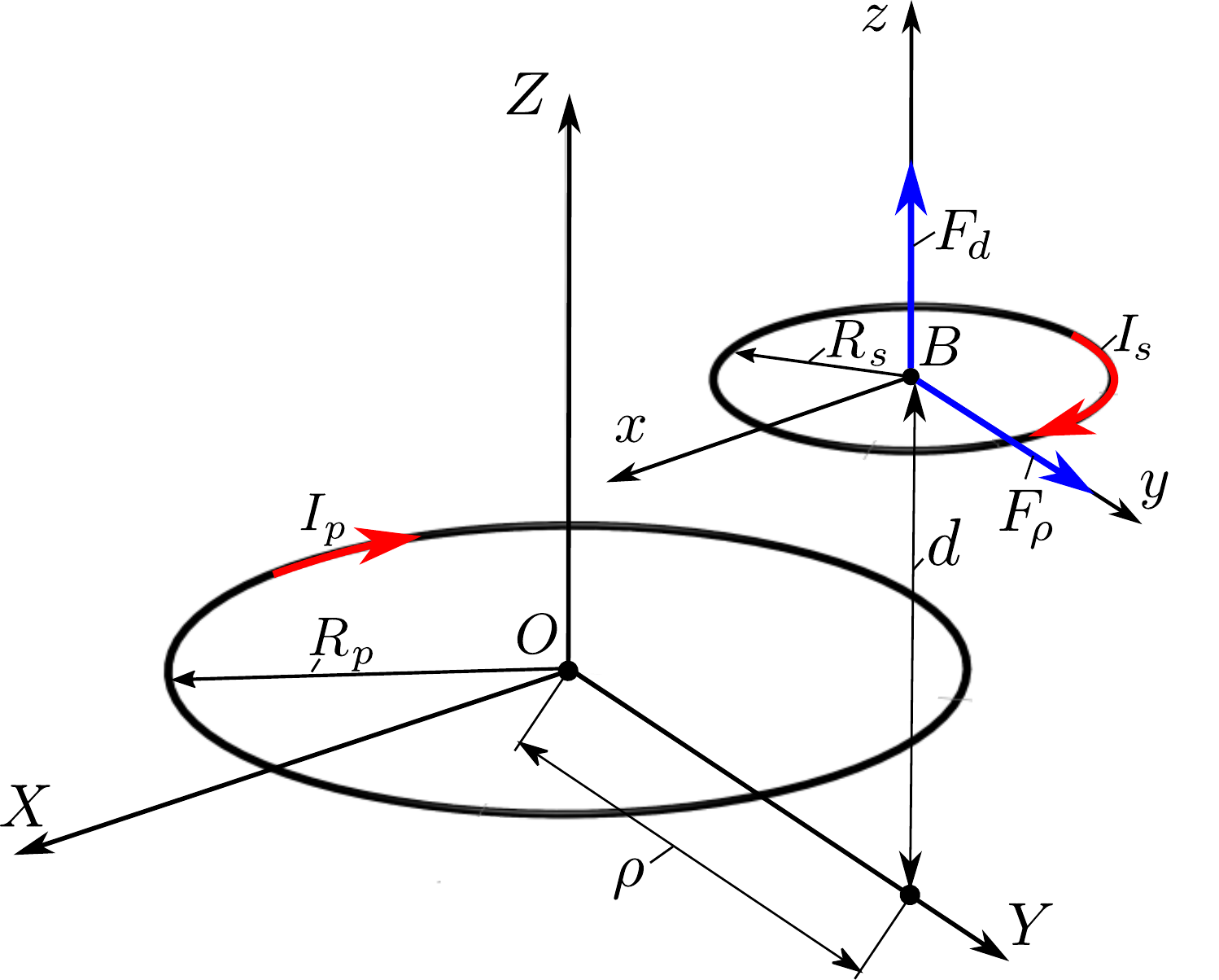}
  \caption{Geometrical scheme  of  circular  filaments with parallel axes denoted via Grover's notation: $\rho$ is the distance between axes; $d$ is the distance between the coils' planes $d=z_B$.   }\label{fig:filaments parallel axes}
\end{figure}
\subsection{Force and torque between circular  filaments with parallel axes}
The scheme for calculation of force and torque between circular  filaments with parallel
axes is shown in Fig. \ref{fig:filaments parallel axes}. The linear misalignment
in the Grover notation can be defined by the geometrical parameter, $d$, which is the distance between the planes of
circles  and the parameter, $\rho$,
is the distance between their axes.
These parameters have the following relationship to the notation defined in this article, namely, $z_B=d$ and $\rho=\sqrt{x_B^2+y_B^2}$. Fig. \ref{fig:filaments parallel axes} shows the particular case, when $\rho=y_B$, which is convenient for calculation of the restoring magnetic force $F_y=F_{\rho}$ and the propulsive magnetic force $F_z=F_d$ corresponding to Example 6 in  Babi\v{c} et al. work \cite{Babic2012a}. Also, the torque $T_{\theta}$ and $T_{\eta}$, which are directed along  the $x$- and $z$-axis, respectively, is  calculated.

\subsubsection*{ Example 1: Force (Example 6, page 74 in Babi\v{c} et al. work \cite{Babic2012a})}\label{sec:exmple1}
In this example, the restoring  force (in Babi\v{c}'s notation, it is denoted as $F_r$) between the primary filament having a radius of $R_p=$\SI{42.5}{\milli\meter} and the secondary one having a radius of $R_s=$\SI{20.0}{\milli\meter} was calculated. The distance between the coils' centres is $\rho=y_B=$\SI{3.0}{\milli\meter}, while the distance, $d=z_B$, between the coils' planes is in a range between 0 and \SI{11}{\milli\meter}. For the calculation both the developed new formula (\ref{eq: x and y first der of MI}) and the one derived by Grover's method (\ref{eq:rho-der GF}) were used.  The results of calculation are summed up in Table \ref{tab:Example1}.
 As it can be seen from  the analysis of Table \ref{tab:Example1} that all results of calculation of restoring force obtained by  both formulas (\ref{eq: x and y first der of MI}) and (\ref{eq:rho-der GF})  are in an excellent agreement with results of calculation performed by  Babi\v{c}'s formula.
\begin{table}[!t]
  \caption{Calculation of the restoring force for Example 1 }\label{tab:Example1}
  \centering
\begin{tabular}{lccc}
 \toprule
  & Babi\v{c}'s&  Grover's method,& This work,\\
$z_B=d$& formula, \cite[Eq. (10)]{Babic2012a},&Eq. (\ref{eq:rho-der GF}),&Eq. (\ref{eq: x and y first der of MI}),\\
 \si{\milli\metre}  & $F_r=F_{\rho}$, \si{\micro\newton}& $F_{\rho}=F_y$, \si{\micro\newton}&$F_y=F_{\rho}$, \si{\micro\newton}\\
  \midrule
  0 & 0.0754774971002899&0.0754774971002898 & 0.0754774971002900\\
 1.0 & 0.0748858332720979&0.0748858332720979& 0.0748858332720979\\
   2.0 &  0.0731367134919867&0.0731367134919868& 0.0731367134919868 \\
   3.0 &  0.0703054618194718&0.0703054618194719& 0.0703054618194720\\
  4.0 & 0.0665103249889932&0.0665103249889937& 0.0665103249889934\\
    5.0 & 0.0619026566955100&0.0619026566955099& 0.0619026566955097\\
      6.0 &  0.0566551686796067&0.0566551686796067&0.0566551686796069\\
       7.0 &  0.0509497255158943&0.0509497255158945&0.0509497255158945\\
        8.0 & 0.0449660177790821&0.0449660177790823& 0.0449660177790822\\
        9.0 & 0.0388721123989230&0.0388721123989233& 0.0388721123989234\\
        10.0& 0.0328174528506593& 0.0328174528506596& 0.0328174528506595\\
       11.0 &0.0269284649078899&0.0269284649078902& 0.0269284649078900\\
        \toprule
\end{tabular}
\end{table}

\subsubsection*{ Example 2: Force (Example 6, page 74 in Babi\v{c} et al. work \cite{Babic2012a})}
The propulsive  force (in Babi\v{c}'s notation, it is denoted as $F_z$) between the primary and second filament under the same  configuration  given in Example 1 was calculated. The results of calculation are shown in Table \ref{tab:Example2}. Analysis of Table \ref{tab:Example2} depicts that all results of calculation obtained by new formula (\ref{eq: z first der of MI}) and formula (\ref{eq:d-der GF}) derived by Grover's method have very good agreement with   results of calculation performed by  Babi\v{c}'s formula.
\begin{table}[!t]
  \caption{Calculation of the propulsive force for Example 2 }\label{tab:Example2}
  \centering
\begin{tabular}{lccc}
 \toprule
  & Babi\v{c}'s&  Grover's method,& This work,\\
$z_B=d$& formula, \cite[Eq. (10)]{Babic2012a},&Eq. (\ref{eq:d-der GF}),&Eq. (\ref{eq: z first der of MI}),\\
 \si{\milli\metre}  & $F_z=F_{d}$, \si{\micro\newton}& $F_{d}=F_z$, \si{\micro\newton}&$F_z=F_{d}$, \si{\micro\newton}\\
  \midrule
  0 & 0.0&0.0 & 0.0\\
 1.0 &  -0.0510570118824195&-0.0510570118824194& -0.0510570118824195\\
   2.0 &  -0.101293579295215&-0.101293579295215&-0.101293579295215 \\
   3.0 &  -0.149926462414138&-0.149926462414137& -0.149926462414138\\
  4.0 & -0.196243385024538&-0.196243385024538&-0.196243385024538\\
    5.0 &  -0.239630444882395&-0.239630444882394& -0.239630444882394\\
      6.0 &   -0.279591228665736&-0.279591228665736&-0.279591228665736\\
       7.0 &  -0.315756838476827&-0.315756838476827&-0.315756838476827\\
        8.0 &  -0.347887153545896&-0.347887153545896& -0.347887153545896\\
        9.0 &  -0.375864540746323&-0.375864540746323& -0.375864540746323\\
        10.0&  -0.399681785157597& -0.399681785157597& -0.399681785157597\\
       11.0 & -0.419426221842137&-0.419426221842137& -0.419426221842137\\
        \toprule
\end{tabular}
\end{table}

\begin{table}[!t]
  \caption{Calculation of the torque $T_{\theta}$ for Example 3 }\label{tab:Example5}
  \centering
\begin{tabular}{lcc}
 \toprule
$z_B=d$  &  Grover's method,& This work,\\
 \si{\milli\metre}  & Eq. (\ref{eq:first psi-der of kernel Grover}), $T_{\theta}$, \si{\nano\newton\metre}&Eq. (\ref{eq: theta and eta first der of MI}), $T_{\theta}$, \si{\nano\newton\metre}\\
  \midrule
  0 & 0 & 0\\
 1.0 &-0.1578740617610107& -0.1578740617610108\\
   2.0 & -0.3116544572018358& -0.3116544572018354 \\
   3.0 &  -0.4575094468248243& -0.4575094468248247\\
  4.0 & -0.5921003456361885& -0.5921003456361873\\
    5.0 & -0.7127567873444469& -0.7127567873444469\\
      6.0 &  -0.8175809895015301&-0.8175809895015308\\
       7.0 &  -0.9054774849564481&-0.905477484956448\\
        8.0 & -0.9761154129579873& -0.9761154129579885\\
        9.0 &-1.029838133074231& -1.029838133074231\\
        10.0& -1.067538737088322& -1.067538737088322\\
       11.0 &-1.090520164249831& -1.090520164249831\\
        \toprule
\end{tabular}
\end{table}

\begin{table}[!t]
  \caption{Calculation of the torque $T_{\eta}$ for Example 3 }\label{tab:Example5a}
  \centering
\begin{tabular}{lcc}
 \toprule
$z_B=d$  & Grover's method,& This work,\\
 \si{\milli\metre} & Eq. (\ref{eq:first psi-der of kernel Grover}), $T_{\psi}$, \si{\nano\newton\metre}&Eq. (\ref{eq: theta and eta first der of MI}), $T_{\eta}$, \si{\nano\newton\metre}\\
  \midrule
  0 & $-1.36317281085437\times10^{-16}$ & 0\\
 1.0 &$-2.000664966411278\times10^{-16}$& 0\\
   2.0 &$-1.781112775806449\times10^{-16}$& 0 \\
   3.0 &  $-1.996595700073574\times10^{-16}$& 0\\
  4.0 & $-1.194158312920886\times10^{-16}$&0\\
    5.0 & $-6.21450714145062\times10^{-17}$& 0\\
      6.0 &  $-9.46986487725448\times10^{-17}$&0\\
       7.0 & $1.10795621370042\times10^{-16}$&0\\
        8.0 & $1.548091424456673\times10^{-16}$& 0\\
        9.0 &$-2.570095735195105\times10^{-16}$& 0\\
        10.0& $8.214737035699823\times10^{-17}$& 0\\
       11.0 &$2.123122223811704\times10^{-16}$& 0\\
        \toprule
\end{tabular}
\end{table}

\subsubsection*{ Example 3: Torque}
Considering the scheme  shown in Fig. \ref{fig:filaments parallel axes},  the torques corresponding the generalized coordinates $\theta$ and $\eta$, namely, $T_{\theta}$  and $T_{\eta}$, respectively, are calculated by using formula (\ref{eq: theta and eta first der of MI}). The calculation was performed for the same coil arrangement as in Example 1. Changing  the distance, $d=z_B$, between the coils' planes in the same range between 0 and \SI{11}{\milli\meter}, the obtained results of calculation of the torque $T_{\theta}$  are depicted in Table \ref{tab:Example5}.
While the results of calculation of the torque $T_{\eta}$  are shown in Table \ref{tab:Example5a}. The analysis of the tables show that the results of calculation are in excellent agreement to each other. Also, worth noting that  calculation of the torque $T_{\eta}$ by means of  Kalantarov-Zeitlin's method provides the zero result for all considered cases, while the Grover method shows small errors.

\begin{figure}[!t]
  \centering
  \includegraphics[width=2.5in]{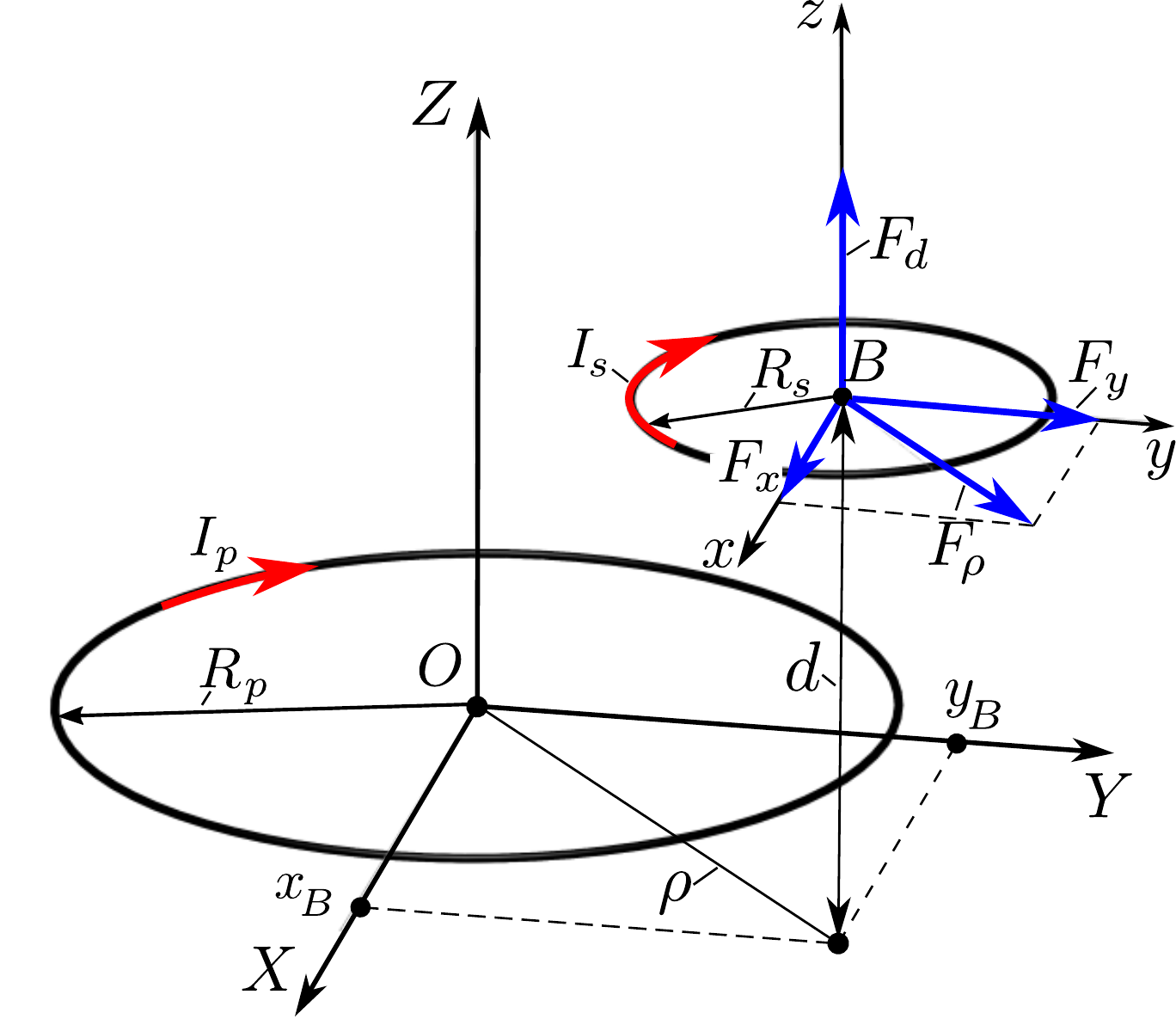}
  \caption{Geometrical scheme  of  circular  filaments with parallel axes corresponding to the general case, when neither  the  $Y$-axis nor $X$-axis coincides with the $\rho$ -axis.   }\label{fig:filaments parallel axes example3}
\end{figure}
\subsubsection*{ Example 4: Force (Example 6, page 74 in Babi\v{c} et al. work \cite{Babic2012a})}
Now, considering the coils having the same radii as in Example 1 and 2, but  let us assume that the centre of secondary coil has the following coordinates: $x_B=y_B=~$\SI{2.1213 }{\milli\meter} and $z_B=~$\SI{8.0}{\milli\meter}. It means that neither  the  $Y$-axis nor $X$-axis coincides with the $\rho$ -axis. In this particular case, the values of coordinates $x_B$  and $y_B$ are corresponding to the Grover parameter $\rho=\sqrt{x_B^2+y_B^2}=~$\SI{3.0 }{\milli\meter}, which has the same value as in the previous examples 1 and 2 (please see Figure \ref{fig:filaments parallel axes example3}).
The restoring   force with its $x$- and $y$-components and propulsive force between the primary and second filament  were calculated. The results are shown in Table \ref{tab:Example3 restoring} and \ref{tab:Example3   propulsive}, respectively.

\begin{table}[!t]
  \caption{Calculation of the restoring force for Example 4 }\label{tab:Example3 restoring}
  \centering
\begin{tabular}{cccc}
 \toprule
  This work,&  $F_x$, \si{\micro\newton}  &$F_y$, \si{\micro\newton}& $\sqrt{F_x^2+F_y^2}$, \si{\micro\newton} \\
     Eq. (\ref{eq: x and y first der of MI})&   $0.031795776094544$ & $0.031795776094544$ & $0.0449660177790823$ \\
    \midrule
 \multicolumn{3}{ l }{ Babi\v{c}'s formula \cite[Eq. (10)]{Babic2012a}, $F_r$, \si{\micro\newton}} & $0.0449660177790821$ \\
   \midrule
    \multicolumn{3}{ l }{ Grover's method, Eq. (\ref{eq:rho-der GF}), $F_{\rho}$, \si{\micro\newton}} &  $0.0449660177790823$ \\
  \toprule
\end{tabular}
\end{table}

\begin{table}[!t]
  \caption{Calculation of the propulsive force for Example 4 }\label{tab:Example3 propulsive}
  \centering
\begin{tabular}{ccc}
 \toprule
    Babi\v{c}'s formula, \cite[Eq. (10)]{Babic2012a},&  Grover's method,& This work,\\
$F_z$, \si{\micro\newton}&Eq. (\ref{eq:d-der GF}), $F_{d}$, \si{\micro\newton} &Eq. (\ref{eq: z first der of MI}), $F_z$, \si{\micro\newton}\\
  \midrule
$-0.347887153545896$&$-0.347887153545896$&$ -0.347887153545896$\\
         \toprule
\end{tabular}
\end{table}

\subsubsection*{ Example 5: Force (Example 8, page 75 in Babi\v{c} et al. work \cite{Babic2012a})}
The primary coil has a radius of \SI{1 }{\meter}, while the secondary of \SI{0.5 }{\meter}. The centre of the secondary coil with respect to the primary one is located at point $B$ having the following coordinate $x_B=y_B=z_B=~$\SI{2.0}{\meter}. The results of calculation are as follows

\vspace*{1.0em}
\begin{tabular}{lccc}
 \toprule
  This work&  $F_x$, Eq. (\ref{eq: x and y first der of MI}) , \si{\nano\newton}  &$F_y$, Eq. (\ref{eq: x and y first der of MI}), \si{\nano\newton}& $F_z$, Eq. (\ref{eq: z first der of MI}), \si{\nano\newton} \\
     &   $-2.745371984357345 $ & $-2.745371984357349$ & $3.509473102444028$ \\
    \midrule
     Babi\v{c}'s&  $F_x$, \si{\nano\newton}  &$F_y$,  \si{\nano\newton}& $F_z$, \si{\nano\newton} \\
   formula  &   $-2.745371984357346 $ & $-2.745371984357346 $ & $3.509473102444032$ \\
    \midrule
  Grover's&  -- &$F_{\rho}$, Eq. (\ref{eq:rho-der GF}), \si{\nano\newton}& $F_d$, Eq. (\ref{eq:d-der GF}), \si{\nano\newton} \\
   method  &   -- & $-3.882542294037291$ & $3.50947310244403 $ \\
  \midrule
     This work,&  $\sqrt{F_x^2+F_y^2}$, \si{\nano\newton}: &  $3.882542294037292$&-- \\
  \toprule
\end{tabular}

\subsubsection*{ Example 6: Force, the special case of $\theta=\pi/2$ (Example 9, page 76 in Babi\v{c} et al. work \cite{Babic2012a})}
The primary coil has a radius of \SI{1 }{\meter}, while the secondary of \SI{0.5 }{\meter}. The centre of the secondary coil with respect to the primary one is located at point $B$ having the following coordinate $x_B=$\SI{1.0}{\meter}, $y_B=$\SI{2.0}{\meter}, $z_B=~$\SI{3.0}{\meter}. The secondary coil is located on the plane $x=$ \SI{1.0}{\meter} ($\theta=\eta=$\SI{\pi/2}{\radian}). For the calculation, Eq. (\ref{eq:first der of SC x and y}) and (\ref{eq:first der of SC z}) are used. Results are

\vspace*{1.0em}
\begin{tabular}{lccc}
 \toprule
  This work&  $F_x$, Eq. (\ref{eq:first der of SC x and y}) , \si{\nano\newton}  &$F_y$, Eq. (\ref{eq:first der of SC x and y}), \si{\nano\newton}& $F_z$, Eq. (\ref{eq:first der of SC z}), \si{\nano\newton} \\
     &   $1.939241379554505 $ & $-1.861181718234279$ & $-2.202382194552672$ \\
    \midrule
    Babi\v{c}'s&  $F_x$, \si{\nano\newton}  &$F_y$,  \si{\nano\newton}& $F_z$, \si{\nano\newton} \\
   formula  &   $1.939241379554508 $ & $-1.861181718234281 $ & $ -2.202382194552672$ \\
  \toprule
\end{tabular}

\subsubsection*{ Example 7: Torque (the  singular case of  Kalantarov-Zeitlin’s method)}
For the same arrangement of coils as in Example 6 above, the torque $T_{\eta}$ is calculated by formula (\ref{eq:first der of SC eta}).
This case corresponds to the singularity of Kalantarov-Zeitlin's method for the calculation of the torque $T_{\theta}$. Due to this fact the value is not available in the table below. The results are

\vspace*{1.0em}
\begin{tabular}{lcc}
 \toprule

 This work, &$T_{\theta}$, \si{\nano\newton\metre}& $T_{\eta}$, \si{\nano\newton\metre} \\
  Eq. (\ref{eq:first der of SC eta}):    & $--$ & $6.860953527497655$ \\
    \midrule
    Grover's method,  &$T_{\theta}$, \si{\nano\newton\metre}& $T_{\psi}$, \si{\nano\newton\metre} \\
   Eq. (\ref{eq:ang coor-der GF}): & $-6.036471731788468$ & $6.860953527497664$\\
  \toprule
\end{tabular}
\vspace*{1.0em}

However, to avoid this difficulty the angle $\theta$ can be chosen enough close to the value $\pi/2$, but not equal to it. Hence, formula (\ref{eq: theta and eta first der of MI}) can be applied. For instance, if the angle $\theta$ is \SI{1.57062}{\radian}, the result of calculation of the  torque $T_{\theta}$ becomes \SI{-6.036471614863915}{\nano\newton\metre} with a relative error of $-1.93696844616845\times10^{-8}$ in comparing with the exact result of calculation obtained by means of Grover's method.

\subsubsection*{ Example 8: Force, the special case of $\theta=\pi/2$ (Example 10, page 76 in Babi\v{c} et al. work \cite{Babic2012a})}
  The primary and secondary coils have the same radii as in example 5. While, the centre of the secondary coil with respect to the primary one is located at point $B$ having the  coordinate: $x_B=y_B=z_B=$\SI{2.0}{\meter}. The secondary coil is located on the plane $y=$ \SI{2.0}{\meter} ($\theta=$ \SI{\pi/2}{\radian} and $\eta=$\SI{0}{\radian}). For the calculation, as in the previous example Eq. (\ref{eq:first der of SC x and y}) and (\ref{eq:first der of SC z}) are used. The results are

\vspace*{1.0em}
\begin{tabular}{lccc}
 \toprule
  This work&  $F_x$, Eq. (\ref{eq:first der of SC x and y}), \si{\nano\newton}  &$F_y$, Eq. (\ref{eq:first der of SC x and y}), \si{\nano\newton}& $F_z$, Eq. (\ref{eq:first der of SC z}), \si{\nano\newton} \\
     &   $4.901398177052338$ & $1.984872313200136$ & $2.582265710169335$ \\
    \midrule
     The Babi\v{c}&  $F_x$, \si{\nano\newton}  &$F_y$,  \si{\nano\newton}& $F_z$, \si{\nano\newton} \\
   formula  &   $4.901398177052345$ & $1.984872313200137 $ & $ 2. 582265710169336$ \\
  \toprule
\end{tabular}

\subsubsection*{ Example 9: Force }
 The primary and secondary coils have the same radii as in example 5 (the primary coil has a radius of \SI{1 }{\meter} and the secondary one of \SI{0.5 }{\meter}).
  The centre of the secondary coil with respect to the primary one is located at point $B$ having the coordinates: $x_B=0$ and $y_B=z_B=$\SI{2.0}{\meter} ($\theta=0$  and $\eta=$\SI{0}{\radian}).
The results of calculation are as follows

\vspace*{1.0em}
\begin{footnotesize}
\begin{tabular}{lcc}
 \toprule
This work:&&\\
   $F_x$, Eq. (\ref{eq: x and y first der of MI}), \si{\nano\newton}
  &$F_y$, Eq. (\ref{eq: x and y first der of MI}), \si{\nano\newton}& $F_z$, Eq. (\ref{eq: z first der of MI}), \si{\nano\newton} \\
      $-2.943923360032078\times10^{-15}$ & $-13.05164071847218$ & $0.5836068102838426$ \\
    \midrule
      The Grover method:  &$F_{\rho}$, Eq. (\ref{eq:rho-der GF}), \si{\nano\newton}& $F_d$, Eq. (\ref{eq:d-der GF}), \si{\nano\newton} \\
    -- & $-13.05164071847219$ & $ 0.583606810283838$\\
  \toprule
\end{tabular}
\end{footnotesize}

\subsubsection*{ Example 10: Force,   the special case of $\theta=\pi/2$}
 The primary and secondary coils have the same radii as in example 5 (the primary coil has a radius of \SI{1 }{\meter} and the secondary one of \SI{0.5 }{\meter}).
  The centre of the secondary coil with respect to the primary one is located at point $B$ having the coordinates: $x_B=0$ and $y_B=z_B=$\SI{2.0}{\meter}. While, the angular orientation of the secondary coil is defined as follows $\theta=\pi/2$  and $\eta=$\SI{0}{\radian}.
The results of calculation are

\vspace*{1.0em}
\begin{footnotesize}
\begin{tabular}{lcc}
 \toprule
This work:&&\\
   $F_x$,  Eq. (\ref{eq:first der of SC x and y}), \si{\nano\newton}
  &$F_y$,  Eq. (\ref{eq:first der of SC x and y}), \si{\nano\newton}& $F_z$, Eq. (\ref{eq:first der of SC z}) , \si{\nano\newton} \\
      $-3.986562883376773\times10^{-16}$ & $9.62480501067982$ & $12.81718822413886$ \\
    \midrule
       The Grover method: &$F_{\rho}$, Eq. (\ref{eq:rho-der GF}), \si{\nano\newton}& $F_d$, Eq. (\ref{eq:d-der GF}), \si{\nano\newton} \\
    -- & $9.624805010679824$ & $ 12.81718822413886$\\
  \toprule
\end{tabular}
\end{footnotesize}

\begin{figure}[!t]
  \centering
  \includegraphics[width=2.5in]{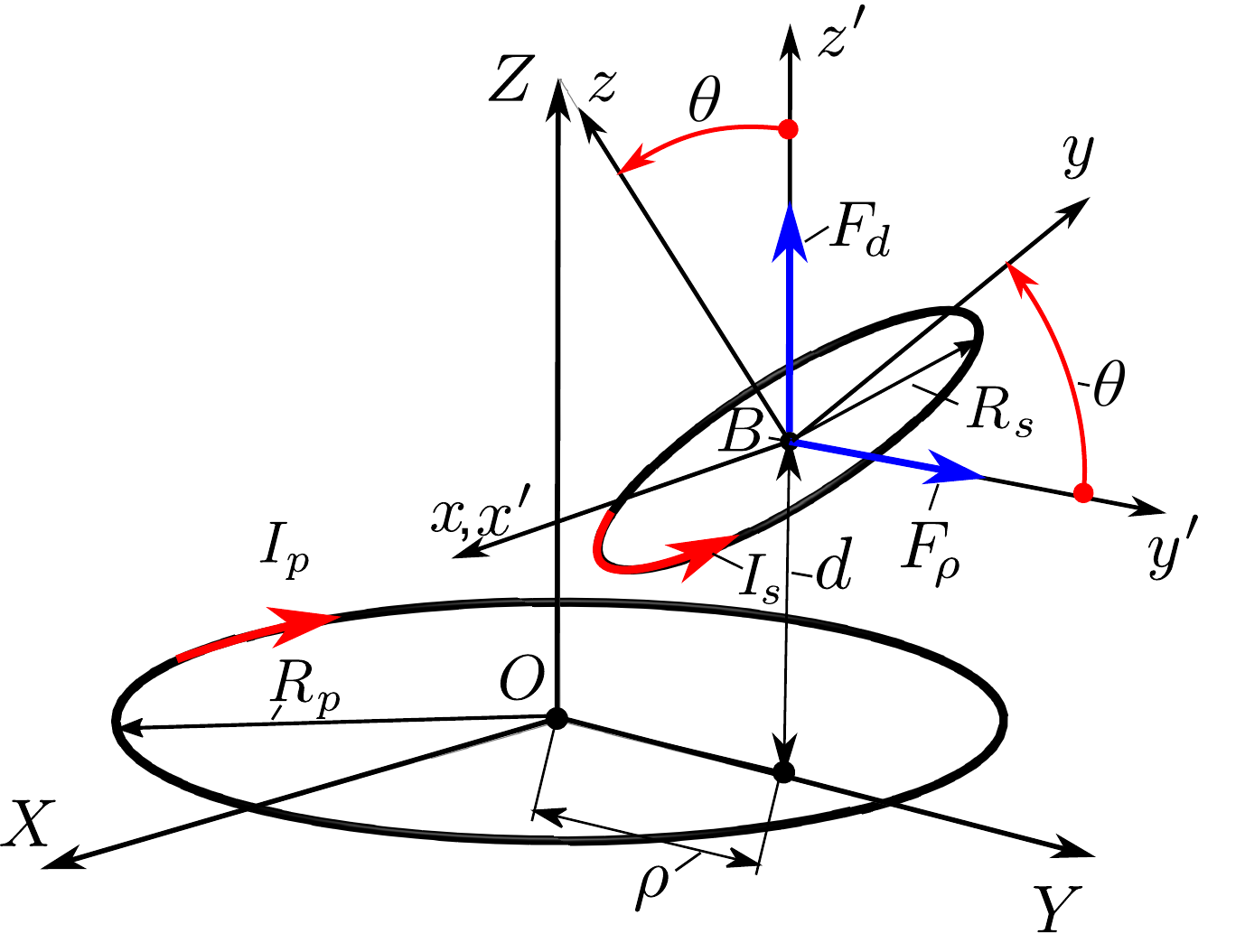}
  \caption{Geometrical scheme  of  circular  filaments with intersect axes.   }\label{fig:filaments intersect axes}
\end{figure}

\subsection{ Force and torque between inclined circular  filaments with intersect axes}

The general scheme of the arrangement of two inclined circular filaments whose axes intersect for calculation of the force and torque is shown in Fig. \ref{fig:filaments intersect axes}. The centre $B$ of the secondary circle is located on the $ZY$- plane and the $xy$ plane of the secondary circle is inclined by the $\theta$ angle ($\eta=0$).

\subsubsection*{ Example 11: Force}
The primary and secondary coils have the same radii as in examples 7 and 8 (the primary coil has a radius of \SI{1 }{\meter} and the secondary one of \SI{0.5 }{\meter}).
Also, similar to Example 7 and 8,  the centre of the secondary coil with respect to the primary one is located at point $B$ having the coordinates: $x_B=0$ and $y_B=z_B=$\SI{2.0}{\meter}, while the angel of $\theta$ is changed in a range from $\pi/12$ to \SI{5\pi/12}{\radian}.
Hence, the results of calculation of force are as follows.

\vspace*{1.0em}
 For the angle of $\theta=$\SI{\pi/12}{\radian}, results are

\begin{footnotesize}
\begin{tabular}{lcc}
 \toprule
This work:&&\\
   $F_x$,  Eq. (\ref{eq: x and y first der of MI}), \si{\nano\newton}
  &$F_y$,  Eq. (\ref{eq: x and y first der of MI}), \si{\nano\newton}& $F_z$, Eq. (\ref{eq: z first der of MI}) , \si{\nano\newton} \\
      $-3.925231146709438\times10^{-15}$ & $-10.83934689818066$ & $4.085335195849623$ \\
    \midrule
      The Grover method: &$F_{\rho}$, Eq. (\ref{eq:rho-der GF}), \si{\nano\newton}& $F_d$, Eq. (\ref{eq:d-der GF}), \si{\nano\newton} \\
    -- & $-10.83934689818066$ & $ 4.08533519584963$\\
  \toprule
\end{tabular}
\end{footnotesize}

\vspace*{1.0em}
 For the angle of $\theta=$\SI{\pi/6}{\radian}, results are

\begin{footnotesize}
\begin{tabular}{lcc}
 \toprule
This work:&&\\
   $F_x$,  Eq. (\ref{eq: x and y first der of MI}), \si{\nano\newton}
  &$F_y$,  Eq. (\ref{eq: x and y first der of MI}), \si{\nano\newton}& $F_z$, Eq. (\ref{eq: z first der of MI}) , \si{\nano\newton} \\
      $-2.943923360032078\times10^{-15}$ & $-7.552692503639927$ & $7.633101747693313 $ \\
    \midrule
       The Grover method: &$F_{\rho}$, Eq. (\ref{eq:rho-der GF}), \si{\nano\newton}& $F_d$, Eq. (\ref{eq:d-der GF}), \si{\nano\newton} \\
    -- & $-7.552692503639934$ & $ 7.633101747693303$\\
  \toprule
\end{tabular}
\end{footnotesize}

\vspace*{1.0em}
 For the angle of $\theta=$\SI{\pi/4}{\radian}, results are

\begin{footnotesize}
\begin{tabular}{lcc}
 \toprule
This work:&&\\
   $F_x$,  Eq. (\ref{eq: x and y first der of MI}), \si{\nano\newton}
  &$F_y$,  Eq. (\ref{eq: x and y first der of MI}), \si{\nano\newton}& $F_z$, Eq. (\ref{eq: z first der of MI}) , \si{\nano\newton} \\
      $-2.943923360032078\times10^{-15}$ & $-3.307376210012321$ & $10.68822709927552 $ \\
    \midrule
      The Grover method:&$F_{\rho}$, Eq. (\ref{eq:rho-der GF}), \si{\nano\newton}& $F_d$, Eq. (\ref{eq:d-der GF}), \si{\nano\newton} \\
    -- & $-3.307376210012323$ & $ 10.6882270992755$\\
  \toprule
\end{tabular}
\end{footnotesize}

\vspace*{1.0em}
 For the angle of $\theta=$\SI{\pi/3}{\radian}, results are

\begin{footnotesize}
\begin{tabular}{lcc}
 \toprule
This work:&&\\
   $F_x$,  Eq. (\ref{eq: x and y first der of MI}), \si{\nano\newton}
  &$F_y$,  Eq. (\ref{eq: x and y first der of MI}), \si{\nano\newton}& $F_z$, Eq. (\ref{eq: z first der of MI}) , \si{\nano\newton} \\
      $-9.813077866773594\times10^{-16}$ & $1.423367390491171$ & $12.68704579579398 $ \\
    \midrule
      The Grover method: &$F_{\rho}$, Eq. (\ref{eq:rho-der GF}), \si{\nano\newton}& $F_d$, Eq. (\ref{eq:d-der GF}), \si{\nano\newton} \\
    -- & $1.42336739049117$ & $ 12.68704579579398$\\
  \toprule
\end{tabular}
\end{footnotesize}

\vspace*{1.0em}
For the angle of $\theta=$\SI{5\pi/12}{\radian}, results are

\begin{footnotesize}
\begin{tabular}{lcc}
 \toprule
This work:&&\\
   $F_x$,  Eq. (\ref{eq: x and y first der of MI}), \si{\nano\newton}
  &$F_y$,  Eq. (\ref{eq: x and y first der of MI}), \si{\nano\newton}& $F_z$, Eq. (\ref{eq: z first der of MI}) , \si{\nano\newton} \\
      $4.906538933386797\times10^{-16}$ & $5.934542693142568$ & $13.34958760362382 $ \\
    \midrule
      The Grover method:&$F_{\rho}$, Eq. (\ref{eq:rho-der GF}), \si{\nano\newton}& $F_d$, Eq. (\ref{eq:d-der GF}), \si{\nano\newton} \\
    -- & $5.934542693142566$ & $ 13.3495876036238$\\
  \toprule
\end{tabular}
\end{footnotesize}

\subsubsection*{ Example 12: Torque}
Now, for the same coil arrangement as in  Example 10 above, the components of the torque  were calculated. The results are as follows.

%
%
%

\vspace*{1.0em}
For the angle of $\theta=$\SI{\pi/12}{\radian}, results are

\begin{footnotesize}
\begin{tabular}{lcc}
 \toprule

 This work, &$T_{\theta}$, \si{\nano\newton\metre}& $T_{\eta}$, \si{\nano\newton\metre} \\
  Eq. (\ref{eq: theta and eta first der of MI}):    & $-16.91560720972092$ & $-1.103971260012029\times10^{-15}$ \\
    \midrule
      The Grover method,  &$T_{\theta}$, \si{\nano\newton\metre}& $T_{\psi}$, \si{\nano\newton\metre} \\
   Eq. (\ref{eq:first psi-der of kernel Grover}): & $-16.91560720972094$ & $3.496810701779198\times10^{-15}$\\
  \toprule
\end{tabular}
\end{footnotesize}

\vspace*{1.0em}
For the angle of $\theta=$\SI{\pi/6}{\radian}, results are

\begin{footnotesize}
\begin{tabular}{lcc}
 \toprule

 This work, &$T_{\theta}$, \si{\nano\newton\metre}& $T_{\eta}$, \si{\nano\newton\metre} \\
  Eq. (\ref{eq: theta and eta first der of MI}):    & $-18.05278139644676$ & $-1.717288626685379\times10^{-15}$ \\
    \midrule
      The Grover method,  &$T_{\theta}$, \si{\nano\newton\metre}& $T_{\psi}$, \si{\nano\newton\metre} \\
   Eq. (\ref{eq:first psi-der of kernel Grover}): & $-18.05278139644677$ & $2.99873918328272\times10^{-16}$\\
  \toprule
\end{tabular}
\end{footnotesize}

\vspace*{1.0em}
For the angle of $\theta=$\SI{\pi/4}{\radian}, results are

\begin{footnotesize}
\begin{tabular}{lcc}
 \toprule

 This work, &$T_{\theta}$, \si{\nano\newton\metre}& $T_{\eta}$, \si{\nano\newton\metre} \\
  Eq. (\ref{eq: theta and eta first der of MI}):    & $-17.23198131459169$ & $4.599880250050123\times10^{-15}$ \\
    \midrule
      The Grover method,  &$T_{\theta}$, \si{\nano\newton\metre}& $T_{\psi}$, \si{\nano\newton\metre} \\
   Eq. (\ref{eq:first psi-der of kernel Grover}): & $-17.23198131459168$ & $-2.386186316760677\times10^{-16}$\\
  \toprule
\end{tabular}
\end{footnotesize}

\vspace*{1.0em}
For the angle of $\theta=$\SI{\pi/3}{\radian}, results are

\begin{footnotesize}
\begin{tabular}{lcc}
 \toprule

 This work, &$T_{\theta}$, \si{\nano\newton\metre}& $T_{\eta}$, \si{\nano\newton\metre} \\
  Eq. (\ref{eq: theta and eta first der of MI}):    & $-14.34174801596063$ & $3.925231146709438\times10^{-15}$ \\
    \midrule
      The Grover method,  &$T_{\theta}$, \si{\nano\newton\metre}& $T_{\psi}$, \si{\nano\newton\metre} \\
   Eq. (\ref{eq:first psi-der of kernel Grover}): & $-14.34174801596063$ & $-1.034553191651082\times10^{-15}$\\
  \toprule
\end{tabular}
\end{footnotesize}

\vspace*{1.0em}
For the angle of $\theta=$\SI{5\pi/12}{\radian}, results are

\begin{footnotesize}
\begin{tabular}{lcc}
 \toprule

 This work, &$T_{\theta}$, \si{\nano\newton\metre}& $T_{\eta}$, \si{\nano\newton\metre} \\
  Eq. (\ref{eq: theta and eta first der of MI}):    & $-10.05276095457351$ & $1.177569344012831\times10^{-14}$ \\
    \midrule
      The Grover method,  &$T_{\theta}$, \si{\nano\newton\metre}& $T_{\psi}$, \si{\nano\newton\metre} \\
   Eq. (\ref{eq:first psi-der of kernel Grover}): & $-10.05276095457353$ & $2.72021517637498\times10^{-16}$\\
  \toprule
\end{tabular}
\end{footnotesize}

\subsection{ Force and torque between  circular  filaments arbitrarily positioned in the space}
The validation of the developed formulas, namely, for force calculation (\ref{eq: x and y first der of MI}), (\ref{eq: z first der of MI}) and for torque calculation (\ref{eq: theta and eta first der of MI})  between  circular  filaments arbitrarily positioned in the space 
as shown in Fig. \ref{fig:in   space} is performed by comparison with the results of calculation obtained by utilizing  Grover's method.  The position  of the secondary coil with respect to the primary one is determined by the linear and angular misalignment, in particular, the angular one is defined by the $\eta$- and $\theta$-angle.

\begin{figure}[!t]
  \centering
  \includegraphics[width=3.2in]{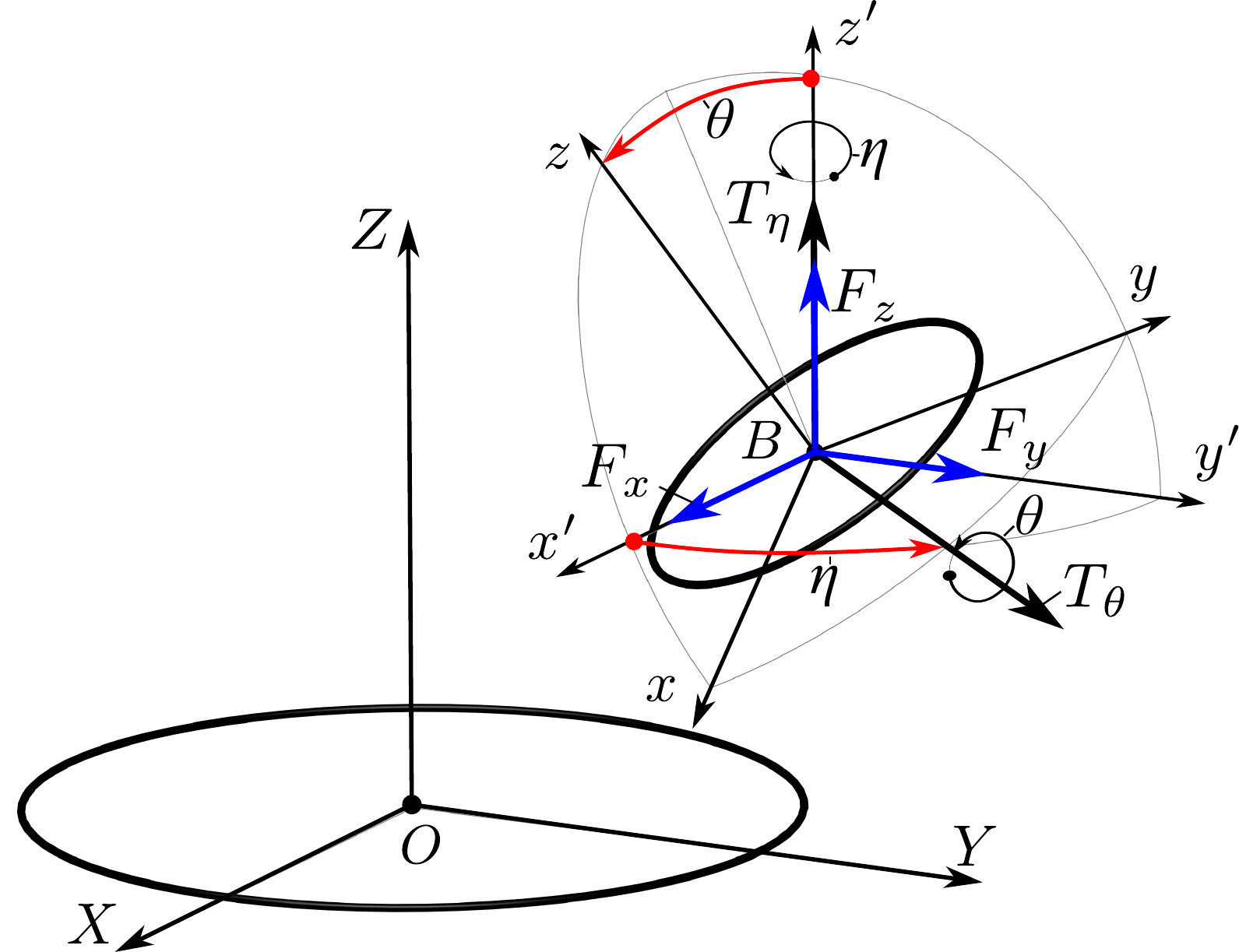}
  \caption{Geometrical scheme  for calculation of force and torque    between  circular  filaments arbitrarily positioned in the space: the angular misalignment is determined by the $\eta$- and $\theta$-angle.   }\label{fig:in space}
\end{figure}

\subsubsection*{ Example 13: Force}
The primary and secondary  circles have  radii $R_p=$\SI{16.0}{\centi\meter} and $R_s=$\SI{10.0}{\centi\meter}, respectively. The centre of the secondary circle is located at $x_B=0$, $y_B=$\SI{4.3301}{\centi\meter}, $z_B=$\SI{17.5}{\centi\meter} and the angle, $\theta$ of \SI{60.0}{\degree}, but the angle $\eta$ is varied in a range from 0 to 360\si{\degree}. The arrangement corresponds to one considered in  Babi\v{c} et al. work \cite{BabicSiroisAkyelEtAl2010} for calculation of mutual inductance.   The results of calculation are summed up below. 

\vspace*{1.0em}
For the angle of $\eta=$\SI{0}{\radian}, results are

\begin{footnotesize}
\begin{tabular}{lcc}
 \toprule
This work:&&\\
   $F_x$,  Eq. (\ref{eq: x and y first der of MI}), \si{\micro\newton}
  &$F_y$,  Eq. (\ref{eq: x and y first der of MI}), \si{\micro\newton}& $F_z$, Eq. (\ref{eq: z first der of MI}) , \si{\nano\newton} \\
      $-9.12816701829942\times10^{-17}$ & $-0.178854052497651$ & $-0.120726650359416$ \\
    \midrule
     The Grover method: &$F_{\rho}$, Eq. (\ref{eq:rho-der GF}), \si{\micro\newton}& $F_d$, Eq. (\ref{eq:d-der GF}), \si{\micro\newton} \\
    -- & $-0.178854052497651$ & $ -0.120726650359416$\\
  \toprule
\end{tabular}
\end{footnotesize}

\vspace*{1.0em}
For the angle of $\eta=$\SI{\pi/6}{\radian}, results are

\begin{footnotesize}
\begin{tabular}{lcc}
 \toprule
This work:&&\\
   $F_x$,  Eq. (\ref{eq: x and y first der of MI}), \si{\micro\newton}
  &$F_y$,  Eq. (\ref{eq: x and y first der of MI}), \si{\micro\newton}& $F_z$, Eq. (\ref{eq: z first der of MI}) , \si{\nano\newton} \\
      $0.0739614721316561$ & $-0.159981513393707$ & $-0.128840188997851$ \\
    \midrule
     The Grover method: &$F_{\rho}$, Eq. (\ref{eq:rho-der GF}), \si{\micro\newton}& $F_d$, Eq. (\ref{eq:d-der GF}), \si{\micro\newton} \\
    -- & $-0.159981513393707$ & $ -0.128840188997851 $\\
  \toprule
\end{tabular}
\end{footnotesize}

\vspace*{1.0em}
For the angle of $\eta=$\SI{\pi/3}{\radian}, results are

\begin{footnotesize}
\begin{tabular}{lcc}
 \toprule
This work:&&\\
   $F_x$,  Eq. (\ref{eq: x and y first der of MI}), \si{\micro\newton}
  &$F_y$,  Eq. (\ref{eq: x and y first der of MI}), \si{\micro\newton}& $F_z$, Eq. (\ref{eq: z first der of MI}) , \si{\nano\newton} \\
      $0.128768354487727$ & $-0.108439678353152$ & $-0.151814417341889$ \\
    \midrule
     The Grover method: &$F_{\rho}$, Eq. (\ref{eq:rho-der GF}), \si{\micro\newton}& $F_d$, Eq. (\ref{eq:d-der GF}), \si{\micro\newton} \\
    -- & $-0.108439678353152$ & $ -0.151814417341889 $\\
  \toprule
\end{tabular}
\end{footnotesize}

\vspace*{1.0em}
For the angle of $\eta=$\SI{\pi/2}{\radian}, results are

\begin{footnotesize}
\begin{tabular}{lcc}
 \toprule
This work:&&\\
   $F_x$,  Eq. (\ref{eq: x and y first der of MI}), \si{\micro\newton}
  &$F_y$,  Eq. (\ref{eq: x and y first der of MI}), \si{\micro\newton}& $F_z$, Eq. (\ref{eq: z first der of MI}) , \si{\nano\newton} \\
      $0.150180118721259$ & $-0.0376756210066879$ & $-0.18544258034698$ \\
    \midrule
     The Grover method: &$F_{\rho}$, Eq. (\ref{eq:rho-der GF}), \si{\micro\newton}& $F_d$, Eq. (\ref{eq:d-der GF}), \si{\micro\newton} \\
    -- & $-0.037675621006688$ & $ -0.18544258034698 $\\
  \toprule
\end{tabular}
\end{footnotesize}

\vspace*{1.0em}
For the angle of $\eta=$\SI{2\pi/3}{\radian}, results are

\begin{footnotesize}
\begin{tabular}{lcc}
 \toprule
This work:&&\\
   $F_x$,  Eq. (\ref{eq: x and y first der of MI}), \si{\micro\newton}
  &$F_y$,  Eq. (\ref{eq: x and y first der of MI}), \si{\micro\newton}& $F_z$, Eq. (\ref{eq: z first der of MI}) , \si{\nano\newton} \\
      $0.13174801806108$ & $0.034039013278426$ & $-0.222277730943632$ \\
    \midrule
     The Grover method: &$F_{\rho}$, Eq. (\ref{eq:rho-der GF}), \si{\micro\newton}& $F_d$, Eq. (\ref{eq:d-der GF}), \si{\micro\newton} \\
    -- & $0.034039013278426$ & $ -0.222277730943632 $\\
  \toprule
\end{tabular}
\end{footnotesize}

\vspace*{1.0em}
For the angle of $\eta=$\SI{5\pi/6}{\radian}, results are

\begin{footnotesize}
\begin{tabular}{lcc}
 \toprule
This work:&&\\
   $F_x$,  Eq. (\ref{eq: x and y first der of MI}), \si{\micro\newton}
  &$F_y$,  Eq. (\ref{eq: x and y first der of MI}), \si{\micro\newton}& $F_z$, Eq. (\ref{eq: z first der of MI}) , \si{\nano\newton} \\
      $0.0768930004431087$ & $0.0873707070288501$ & $-0.251739105122414 $ \\
    \midrule
     The Grover method: &$F_{\rho}$, Eq. (\ref{eq:rho-der GF}), \si{\micro\newton}& $F_d$, Eq. (\ref{eq:d-der GF}), \si{\micro\newton} \\
    -- & $0.0873707070288498$ & $ -0.251739105122414 $\\
  \toprule
\end{tabular}
\end{footnotesize}

\vspace*{1.0em}
For the angle of $\eta=$\SI{\pi}{\radian}, results are

\begin{footnotesize}
\begin{tabular}{lcc}
 \toprule
This work:&&\\
   $F_x$,  Eq. (\ref{eq: x and y first der of MI}), \si{\micro\newton}
  &$F_y$,  Eq. (\ref{eq: x and y first der of MI}), \si{\micro\newton}& $F_z$, Eq. (\ref{eq: z first der of MI}) , \si{\nano\newton} \\
      $-2.38736675863216\times10^{-16}$ & $0.107078490953807$ & $-0.263132411165356 $ \\
    \midrule
     The Grover method: &$F_{\rho}$, Eq. (\ref{eq:rho-der GF}), \si{\micro\newton}& $F_d$, Eq. (\ref{eq:d-der GF}), \si{\micro\newton} \\
    -- & $0.107078490953807$ & $ -0.263132411165356$\\
  \toprule
\end{tabular}
\end{footnotesize}

\vspace*{1.0em}
For the angle of $\eta=$\SI{7\pi/6}{\radian}, results are

\begin{footnotesize}
\begin{tabular}{lcc}
 \toprule
This work:&&\\
   $F_x$,  Eq. (\ref{eq: x and y first der of MI}), \si{\micro\newton}
  &$F_y$,  Eq. (\ref{eq: x and y first der of MI}), \si{\micro\newton}& $F_z$, Eq. (\ref{eq: z first der of MI}) , \si{\nano\newton} \\
      $-0.0768930004431088$ & $0.08737070702885$ & $-0.251739105122414 $ \\
    \midrule
     The Grover method: &$F_{\rho}$, Eq. (\ref{eq:rho-der GF}), \si{\micro\newton}& $F_d$, Eq. (\ref{eq:d-der GF}), \si{\micro\newton} \\
    -- & $0.0873707070288499$ & $ -0.251739105122414$\\
  \toprule
\end{tabular}
\end{footnotesize}

\vspace*{1.0em}
For the angle of $\eta=$\SI{4\pi/3}{\radian}, results are

\begin{footnotesize}
\begin{tabular}{lcc}
 \toprule
This work:&&\\
   $F_x$,  Eq. (\ref{eq: x and y first der of MI}), \si{\micro\newton}
  &$F_y$,  Eq. (\ref{eq: x and y first der of MI}), \si{\micro\newton}& $F_z$, Eq. (\ref{eq: z first der of MI}) , \si{\nano\newton} \\
      $-0.13174801806108$ & $0.0340390132784261$ & $-0.222277730943632 $ \\
    \midrule
     The Grover method: &$F_{\rho}$, Eq. (\ref{eq:rho-der GF}), \si{\micro\newton}& $F_d$, Eq. (\ref{eq:d-der GF}), \si{\micro\newton} \\
    -- & $0.0340390132784261$ & $ -0.222277730943632$\\
  \toprule
\end{tabular}
\end{footnotesize}

\vspace*{1.0em}
For the angle of $\eta=$\SI{3\pi/2}{\radian}, results are

\begin{footnotesize}
\begin{tabular}{lcc}
 \toprule
This work:&&\\
   $F_x$,  Eq. (\ref{eq: x and y first der of MI}), \si{\micro\newton}
  &$F_y$,  Eq. (\ref{eq: x and y first der of MI}), \si{\micro\newton}& $F_z$, Eq. (\ref{eq: z first der of MI}) , \si{\nano\newton} \\
      $-0.150180118721259$ & $-0.0376756210066879$ & $-0.18544258034698 $ \\
    \midrule
     The Grover method: &$F_{\rho}$, Eq. (\ref{eq:rho-der GF}), \si{\micro\newton}& $F_d$, Eq. (\ref{eq:d-der GF}), \si{\micro\newton} \\
    -- & $-0.0376756210066879$ & $-0.18544258034698$\\
  \toprule
\end{tabular}
\end{footnotesize}

\vspace*{1.0em}
For the angle of $\eta=$\SI{5\pi/3}{\radian}, results are

\begin{footnotesize}
\begin{tabular}{lcc}
 \toprule
This work:&&\\
   $F_x$,  Eq. (\ref{eq: x and y first der of MI}), \si{\micro\newton}
  &$F_y$,  Eq. (\ref{eq: x and y first der of MI}), \si{\micro\newton}& $F_z$, Eq. (\ref{eq: z first der of MI}) , \si{\nano\newton} \\
      $-0.128768354487727$ & $-0.108439678353152$ & $-0.151814417341889 $ \\
    \midrule
     The Grover method: &$F_{\rho}$, Eq. (\ref{eq:rho-der GF}), \si{\micro\newton}& $F_d$, Eq. (\ref{eq:d-der GF}), \si{\micro\newton} \\
    -- & $-0.108439678353152$ & $-0.151814417341889$\\
  \toprule
\end{tabular}
\end{footnotesize}

\vspace*{1.0em}
For the angle of $\eta=$\SI{11\pi/6}{\radian}, results are

\begin{footnotesize}
\begin{tabular}{lcc}
 \toprule
This work:&&\\
   $F_x$,  Eq. (\ref{eq: x and y first der of MI}), \si{\micro\newton}
  &$F_y$,  Eq. (\ref{eq: x and y first der of MI}), \si{\micro\newton}& $F_z$, Eq. (\ref{eq: z first der of MI}) , \si{\nano\newton} \\
      $-0.0739614721316564$ & $-0.159981513393707$ & $-0.128840188997851 $ \\
    \midrule
     The Grover method: &$F_{\rho}$, Eq. (\ref{eq:rho-der GF}), \si{\micro\newton}& $F_d$, Eq. (\ref{eq:d-der GF}), \si{\micro\newton} \\
    -- & $-0.159981513393707$ & $-0.128840188997851$\\
  \toprule
\end{tabular}
\end{footnotesize}

\vspace*{1.0em}
For the angle of $\eta=$\SI{2\pi}{\radian}, results are

\begin{footnotesize}
\begin{tabular}{lcc}
 \toprule
This work:&&\\
   $F_x$,  Eq. (\ref{eq: x and y first der of MI}), \si{\micro\newton}
  &$F_y$,  Eq. (\ref{eq: x and y first der of MI}), \si{\micro\newton}& $F_z$, Eq. (\ref{eq: z first der of MI}) , \si{\nano\newton} \\
     $-2.24693341988909\times10^{-16}$ & $-0.178854052497651$ & $-0.120726650359416 $ \\
    \midrule
     The Grover method: &$F_{\rho}$, Eq. (\ref{eq:rho-der GF}), \si{\micro\newton}& $F_d$, Eq. (\ref{eq:d-der GF}), \si{\micro\newton} \\
    -- & $-0.178854052497651$ & $-0.120726650359416$\\
  \toprule
\end{tabular}
\end{footnotesize}

\subsubsection*{ Example 14: Torque}
For the same arrangement of coils as in Example 13 above, the torque is calculated.
Results are shown below.

\vspace*{1.0em}
For the angle of $\theta=$\SI{0}{\radian}, results are

\begin{footnotesize}
\begin{tabular}{lcc}
 \toprule

 This work, &$T_{\theta}$, \si{\nano\newton\metre}& $T_{\eta}$, \si{\nano\newton\metre} \\
  Eq. (\ref{eq: theta and eta first der of MI}):    & $-35.67279850151469$ & $-2.10650008114602\times10^{-15}$ \\
    \midrule
      The Grover method,  &$T_{\theta}$, \si{\nano\newton\metre}& $T_{\psi}$, \si{\nano\newton\metre} \\
   Eq. (\ref{eq:first psi-der of kernel Grover}): & $-35.67279850151468$ & $7.820828205289753\times10^{-16}$\\
  \toprule
\end{tabular}
\end{footnotesize}

\vspace*{1.0em}
For the angle of $\eta=$\SI{\pi/6}{\radian}, results are

\begin{footnotesize}
\begin{tabular}{lcc}
 \toprule

 This work, &$T_{\theta}$, \si{\nano\newton\metre}& $T_{\eta}$, \si{\nano\newton\metre} \\
  Eq. (\ref{eq: theta and eta first der of MI}):    & $-34.95015525890449$ & $3.202605704772845$ \\
    \midrule
      The Grover method,  &$T_{\theta}$, \si{\nano\newton\metre}& $T_{\psi}$, \si{\nano\newton\metre} \\
   Eq. (\ref{eq:first psi-der of kernel Grover}): & $-34.95015525890452$ & $3.202605704772839$\\
  \toprule
\end{tabular}
\end{footnotesize}

\vspace*{1.0em}
For the angle of $\eta=$\SI{\pi/3}{\radian}, results are

\begin{footnotesize}
\begin{tabular}{lcc}
 \toprule

 This work, &$T_{\theta}$, \si{\nano\newton\metre}& $T_{\eta}$, \si{\nano\newton\metre} \\
  Eq. (\ref{eq: theta and eta first der of MI}):    & $-32.95959317516494$ & $5.575798517673081$ \\
    \midrule
      The Grover method,  &$T_{\theta}$, \si{\nano\newton\metre}& $T_{\psi}$, \si{\nano\newton\metre} \\
   Eq. (\ref{eq:first psi-der of kernel Grover}): & $-32.95959317516493$ & $5.575798517673075$\\
  \toprule
\end{tabular}
\end{footnotesize}

\vspace*{1.0em}
For the angle of $\eta=$\SI{\pi/2}{\radian}, results are

\begin{footnotesize}
\begin{tabular}{lcc}
 \toprule

 This work, &$T_{\theta}$, \si{\nano\newton\metre}& $T_{\eta}$, \si{\nano\newton\metre} \\
  Eq. (\ref{eq: theta and eta first der of MI}):    & $-30.16106072467219$ & $6.502949320749249$ \\
    \midrule
      The Grover method,  &$T_{\theta}$, \si{\nano\newton\metre}& $T_{\psi}$, \si{\nano\newton\metre} \\
   Eq. (\ref{eq:first psi-der of kernel Grover}): & $-30.16106072467218$ & $6.502949320749247$\\
  \toprule
\end{tabular}
\end{footnotesize}

\vspace*{1.0em}
For the angle of $\eta=$\SI{2\pi/3}{\radian}, results are

\begin{footnotesize}
\begin{tabular}{lcc}
 \toprule

 This work, &$T_{\theta}$, \si{\nano\newton\metre}& $T_{\eta}$, \si{\nano\newton\metre} \\
  Eq. (\ref{eq: theta and eta first der of MI}):    & $-27.19599671483476$ & $5.704820930062833$ \\
    \midrule
      The Grover method,  &$T_{\theta}$, \si{\nano\newton\metre}& $T_{\psi}$, \si{\nano\newton\metre} \\
   Eq. (\ref{eq:first psi-der of kernel Grover}): & $-27.19599671483475$ & $5.704820930062828$\\
  \toprule
\end{tabular}
\end{footnotesize}

\vspace*{1.0em}
For the angle of $\eta=$\SI{5\pi/6}{\radian}, results are

\begin{footnotesize}
\begin{tabular}{lcc}
 \toprule

 This work, &$T_{\theta}$, \si{\nano\newton\metre}& $T_{\eta}$, \si{\nano\newton\metre} \\
  Eq. (\ref{eq: theta and eta first der of MI}):    & $-24.8616946084914$ & $3.329543812187064$ \\
    \midrule
      The Grover method,  &$T_{\theta}$, \si{\nano\newton\metre}& $T_{\psi}$, \si{\nano\newton\metre} \\
   Eq. (\ref{eq:first psi-der of kernel Grover}): & $-24.86169460849138$ & $3.329543812187052$\\
  \toprule
\end{tabular}
\end{footnotesize}

\vspace*{1.0em}
For the angle of $\eta=$\SI{\pi}{\radian}, results are

\begin{footnotesize}
\begin{tabular}{lcc}
 \toprule

 This work, &$T_{\theta}$, \si{\nano\newton\metre}& $T_{\eta}$, \si{\nano\newton\metre} \\
  Eq. (\ref{eq: theta and eta first der of MI}):    & $-23.96174184275239$ & $5.617333549722721\times10^{-15}$ \\
    \midrule
      The Grover method,  &$T_{\theta}$, \si{\nano\newton\metre}& $T_{\psi}$, \si{\nano\newton\metre} \\
   Eq. (\ref{eq:first psi-der of kernel Grover}): & $-23.96174184275239$ & $4.791446029057778\times10^{-15}$\\
  \toprule
\end{tabular}
\end{footnotesize}

\vspace*{1.0em}
For the angle of $\eta=$\SI{7\pi/6}{\radian}, results are

\begin{footnotesize}
\begin{tabular}{lcc}
 \toprule

 This work, &$T_{\theta}$, \si{\nano\newton\metre}& $T_{\eta}$, \si{\nano\newton\metre} \\
  Eq. (\ref{eq: theta and eta first der of MI}):    & $-24.8616946084914$ & $-3.329543812187048$ \\
    \midrule
      The Grover method,  &$T_{\theta}$, \si{\nano\newton\metre}& $T_{\psi}$, \si{\nano\newton\metre} \\
   Eq. (\ref{eq:first psi-der of kernel Grover}): & $-24.86169460849138$ & $-3.329543812187046$\\
  \toprule
\end{tabular}
\end{footnotesize}

\vspace*{1.0em}
For the angle of $\eta=$\SI{4\pi/3}{\radian}, results are

\begin{footnotesize}
\begin{tabular}{lcc}
 \toprule

 This work, &$T_{\theta}$, \si{\nano\newton\metre}& $T_{\eta}$, \si{\nano\newton\metre} \\
  Eq. (\ref{eq: theta and eta first der of MI}):    & $-27.19599671483477$ & $-5.704820930062826$ \\
    \midrule
      The Grover method,  &$T_{\theta}$, \si{\nano\newton\metre}& $T_{\psi}$, \si{\nano\newton\metre} \\
   Eq. (\ref{eq:first psi-der of kernel Grover}): & $-27.19599671483473$ & $-5.704820930062823$\\
  \toprule
\end{tabular}
\end{footnotesize}

\vspace*{1.0em}
For the angle of $\eta=$\SI{3\pi/2}{\radian}, results are

\begin{footnotesize}
\begin{tabular}{lcc}
 \toprule

 This work, &$T_{\theta}$, \si{\nano\newton\metre}& $T_{\eta}$, \si{\nano\newton\metre} \\
  Eq. (\ref{eq: theta and eta first der of MI}):    & $-30.16106072467218$ & $-6.502949320749249$ \\
    \midrule
      The Grover method,  &$T_{\theta}$, \si{\nano\newton\metre}& $T_{\psi}$, \si{\nano\newton\metre} \\
   Eq. (\ref{eq:first psi-der of kernel Grover}): & $-30.16106072467218$ & $-6.502949320749249$\\
  \toprule
\end{tabular}
\end{footnotesize}

\vspace*{1.0em}
For the angle of $\eta=$\SI{5\pi/3}{\radian}, results are

\begin{footnotesize}
\begin{tabular}{lcc}
 \toprule

 This work, &$T_{\theta}$, \si{\nano\newton\metre}& $T_{\eta}$, \si{\nano\newton\metre} \\
  Eq. (\ref{eq: theta and eta first der of MI}):    & $-32.95959317516493$ & $-5.575798517673075$ \\
    \midrule
      The Grover method,  &$T_{\theta}$, \si{\nano\newton\metre}& $T_{\psi}$, \si{\nano\newton\metre} \\
   Eq. (\ref{eq:first psi-der of kernel Grover}): & $-32.95959317516497$ & $-5.575798517673072$\\
  \toprule
\end{tabular}
\end{footnotesize}

\vspace*{1.0em}
For the angle of $\eta=$\SI{11\pi/6}{\radian}, results are

\begin{footnotesize}
\begin{tabular}{lcc}
 \toprule

 This work, &$T_{\theta}$, \si{\nano\newton\metre}& $T_{\eta}$, \si{\nano\newton\metre} \\
  Eq. (\ref{eq: theta and eta first der of MI}):    & $-34.95015525890449$ & $-3.202605704772844$ \\
    \midrule
      The Grover method,  &$T_{\theta}$, \si{\nano\newton\metre}& $T_{\psi}$, \si{\nano\newton\metre} \\
   Eq. (\ref{eq:first psi-der of kernel Grover}): & $-34.95015525890452$ & $-3.202605704772843$\\
  \toprule
\end{tabular}
\end{footnotesize}

\vspace*{1.0em}
For the angle of $\eta=$\SI{2\pi}{\radian}, results are

\begin{footnotesize}
\begin{tabular}{lcc}
 \toprule

 This work, &$T_{\theta}$, \si{\nano\newton\metre}& $T_{\eta}$, \si{\nano\newton\metre} \\
  Eq. (\ref{eq: theta and eta first der of MI}):    & $-35.6727985015147$ & $-1.40433338743068\times10^{-15}$ \\
    \midrule
      The Grover method,  &$T_{\theta}$, \si{\nano\newton\metre}& $T_{\psi}$, \si{\nano\newton\metre} \\
   Eq. (\ref{eq:first psi-der of kernel Grover}): & $-35.6727985015147$ & $3.112983652841607\times10^{-16}$\\
  \toprule
\end{tabular}
\end{footnotesize}

\section{Conclusion}

We derived new formulas, namely, (\ref{eq: x and y first der of MI}), (\ref{eq: z first der of MI}), (\ref{eq: theta and eta first der of MI}), (\ref{eq:first der of SC x and y}), (\ref{eq:first der of SC z}) and (\ref{eq:first der of SC eta})   for calculation of force and torque between two circular filaments arbitrarily oriented in space presented in the integral analytical form,  whose kernel function is expressed in terms of the elliptic integrals of the
first and second kinds. 
In particular, formulas (\ref{eq:first der of SC x and y}), (\ref{eq:first der of SC z}) and (\ref{eq:first der of SC eta}) are applied for the special case ($\theta=$\SI{\pi/2}{\radian}), when the circular filaments are mutually perpendicular to each other.  For calculation of the torque $T_{\theta}$, this special case corresponds to the singularity one.
To avoid this difficulty, it is suggested that the angle $\theta$ can be chosen enough close to
the value $\pi/2$, but not equal to it and calculation is performed by means of formula (\ref{eq: theta and eta first der of MI}). Thus, the developed formulas are applicable for all possible arrangements between two circular filaments.

 New
developed formulas have been successfully validated through a number of
examples available in the literature and direct comparison with results of calculation of force and torque performed by expressions derived by   Grover's  method.
Besides, the obtained formulas can be easily programmed, they are
intuitively understandable for application.

\section*{Acknowledgment}
Kirill Poletkin acknowledges with thanks the support from German Research Foundation (Grant KO 1883/37-1) under the priority programme SPP 2206.

%
\appendix

\section{Calculation of force and torque between two arbitrarily
oriented circular filaments using Grover’s formula of mutual inductance \cite[page 207, Eq. (179)]{Grover2004}}
\label{app:Grover}

\begin{figure}[!t]
  \centering
  \includegraphics[width=2.5in]{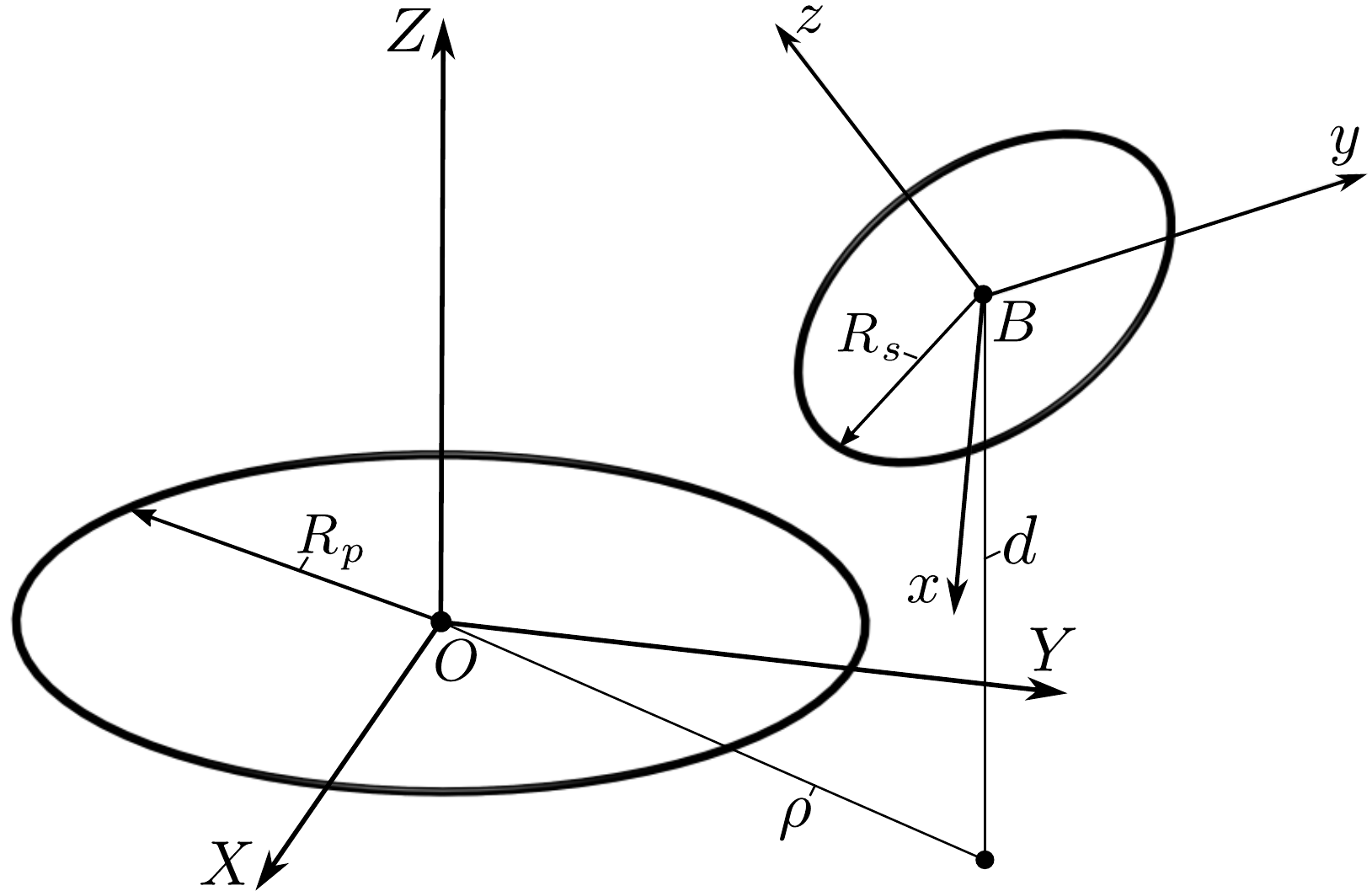}
  \caption{Geometrical scheme of arbitrarily oriented circular filaments: the Grover notations.   }\label{fig:filaments Grover notation}
\end{figure}

According to Grover's notations, the linear misalignment of the centre of the secondary circle is characterised by two parameters, namely, $d=z_B$ and $\rho=\sqrt{x_B^2+y_B^2}$ as shown in Figure \ref{fig:filaments Grover notation}. Besides that the angular misalignment is defined in accordance with the first manner as shown in Fig. \ref{fig:Grover angles}, but keeping the original Grover's notation the angle, $\eta$, is replaced by  $\psi$. In absence of the angular misalignment, the CF $xyz$ assigned to the secondary circle is oriented  in the following way. The $z$-axis is directed upward along the $d$-line, while the $y$-axis is parallel to the $\rho$-line and directed in continuation of the $\rho$-line. Then adopting the above considered notations, Grover's formula for calculation of mutual inductance between two circular filaments can be written as
\begin{equation}\label{eq:GROVER FORMULA}
  M=\frac{\mu_0\sqrt{R_pR_s}}{2\pi}\int_{0}^{2\pi} U\cdot\Psi(k)d\varphi,
\end{equation}
where
\begin{equation}\label{eq:U Grover}
  U=U({\gamma},\theta,\psi)=\frac{R({\gamma},\theta,\psi)}{{V}^{1.5}}=\frac{\cos\theta-\gamma(\cos\psi\cos\varphi-\sin\psi\cos\theta\sin\varphi)}{{V}^{1.5}},
\end{equation}
\begin{equation}\label{eq:V Grover}
V=V({\gamma},\theta,\psi)=\sqrt{1-{\cos(\varphi)}^2{\sin(\theta)}^2+2\gamma(\sin\psi\sin\varphi-\cos\varphi\cos\psi\cos\theta)+\gamma^2},
\end{equation}
\begin{equation}\label{eq:Psi Grover}
   \Psi(k)=\frac{2}{k}\left[\left(1-\frac{k^2}{2}\right)K(k)-E(k)\right],
\end{equation}
\begin{equation}\label{eq:k Grover}
\begin{array}{l}
   {\displaystyle k^2=k^2({\gamma},\Delta,\theta,\psi)=\frac{4\alpha{V}}{(\alpha{V}+1)^2+z^2},
}\\
 {\displaystyle \alpha=R_s/R_p,\;   \Delta=d/R_p,\; \gamma=\rho/R_s},\; z={\Delta}-{\alpha}\sin\theta\cos\varphi.
\end{array}
\end{equation}
The kernel of formula (\ref{eq:GROVER FORMULA}) is
\begin{equation}\label{eq:kernel NF}
  \mathrm{Kr}= U\cdot\Psi(k).
\end{equation}
Accounting for (\ref{eq:U Grover}), (\ref{eq:V Grover}), (\ref{eq:Psi Grover}) and (\ref{eq:k Grover}), the $\rho$-derivative of the kernel becomes
\begin{equation}\label{eq:first rho-der of kernel Grover}
  {\displaystyle  \frac{\partial\mathrm{Kr}}{\partial \rho}=\frac{\partial\mathrm{Kr}}{\partial \gamma}\frac{1}{R_s}=\frac{1}{R_s}\cdot\left[\frac{\partial U}{\partial \gamma} \cdot\Psi(k)+U\cdot\frac{d \Psi(k)}{d k} \cdot\frac{\partial k}{\partial \gamma}\right],}
\end{equation}
where
\begin{equation}\label{eq:dU-gamma Grover}
 \begin{array}{l}
  {\displaystyle  \frac{\partial U}{\partial \gamma}=\left({\displaystyle\frac{\partial R}{\partial \gamma}}\cdot V-1.5\cdot R\cdot\frac{\partial V}{\partial \gamma}\right)\bigg/{{V}^{2.5}},}\\
     {\displaystyle \frac{\partial R}{\partial \gamma}=-(\cos\psi\cos\varphi-\sin\psi\cos\theta\sin\varphi),
}\\
   {\displaystyle \frac{\partial V}{\partial \gamma}=\frac{\sin\psi\,\sin\varphi-\cos\varphi\,\cos\psi\,\cos\theta+\gamma}{V},
}\\
   {\displaystyle \frac{\partial k}{\partial \gamma}=\frac{2/k-k(\alpha{V}+1)}{(\alpha{V}+1)^2+{z}^2}\cdot\alpha\frac{\partial V}{\partial \gamma},
}
   \end{array}
\end{equation}
\begin{equation}\label{eq:dPsi Grover}
 \frac{d \Psi(k)}{d k}=\frac{2}{k^2}\left[\frac{2-k^2}{2(1-k^2)}E(k)-K(k)\right].
\end{equation}
The $d$-derivative of the kernel is
\begin{equation}\label{eq:first d-der of kernel Grover}
  {\displaystyle  \frac{\partial\mathrm{Kr}}{\partial d}=\frac{\partial\mathrm{Kr}}{\partial \Delta}\frac{1}{R_p}=\frac{1}{R_p}\cdot U\cdot\frac{d \Psi(k)}{d k} \cdot\frac{\partial k}{\partial \Delta},}
\end{equation}
where
\begin{equation}\label{eq:dk-d Grover}
 \begin{array}{l}
   {\displaystyle \frac{\partial k}{\partial \Delta}=-\frac{k\cdot z }{(\alpha{V}+1)^2+{z}^2}\cdot\frac{\partial z}{\partial \Delta},
}\\
  {\displaystyle \frac{\partial z}{\partial \Delta}=1.
}
   \end{array}
\end{equation}
Note that in Eq. (\ref{eq:first d-der of kernel Grover}) the $k$-derivative of $\Psi(k)$ is defined  similarly as in Eq. (\ref{eq:dPsi Grover}).  The derivatives of the kernel with respect to the angular coordinates are as follows. The $\theta$-derivative is
\begin{equation}\label{eq:first theta-der of kernel Grover}
  {\displaystyle  \frac{\partial\mathrm{Kr}}{\partial \theta}=\frac{\partial U}{\partial \theta} \cdot\Psi(k)+U\cdot\frac{d \Psi(k)}{d k} \cdot\frac{\partial k}{\partial \theta},}
\end{equation}
where
\begin{equation}\label{eq:dU- and dR- and dk-theta Grover}
 \begin{array}{l}
  {\displaystyle  \frac{\partial U}{\partial \theta}=\left({\displaystyle\frac{\partial R}{\partial \theta}}\cdot V-1.5\cdot R\cdot\frac{\partial V}{\partial \theta}\right)\bigg/{{V}^{2.5}},}\\
     {\displaystyle \frac{\partial R}{\partial \theta}=-\sin\theta\cdot(1+\gamma\cdot\sin\varphi\,\sin\psi),
}\\
   {\displaystyle \frac{\partial V}{\partial \theta}=\frac{-\sin\theta\cdot({\cos(\varphi)}^2\,\cos\theta -\gamma\cdot\cos\varphi\,\cos\psi)}{V},
}\\
   {\displaystyle \frac{\partial k}{\partial \theta}=\frac{\left[2/k-k(\alpha{V}+1)\right]{\displaystyle \cdot\alpha\frac{\partial V}{\partial \theta} }-k\cdot z \cdot{\displaystyle \frac{\partial z}{\partial \theta}} }{(\alpha{V}+1)^2+{z}^2},
}\\ {\displaystyle \frac{\partial z}{\partial \theta}=-\alpha\cdot\cos\theta\cos\varphi.
}
   \end{array}
\end{equation}
The $\psi$-derivative is
\begin{equation}\label{eq:first psi-der of kernel Grover}
  {\displaystyle  \frac{\partial\mathrm{Kr}}{\partial \psi}=\frac{\partial U}{\partial \psi} \cdot\Psi(k)+U\cdot\frac{d \Psi(k)}{d k} \cdot\frac{\partial k}{\partial \psi},}
\end{equation}
where
\begin{equation}\label{eq:dU- and dR- and dk-theta Grover}
 \begin{array}{l}
  {\displaystyle  \frac{\partial U}{\partial \psi}=\left({\displaystyle\frac{\partial R}{\partial \psi}}\cdot V-1.5\cdot R\cdot\frac{\partial V}{\partial \psi}\right)\bigg/{{V}^{2.5}},}\\
     {\displaystyle \frac{\partial R}{\partial \psi}=\gamma\cdot(\cos\varphi\,\sin\psi+\sin\varphi\,\cos\psi\,\cos\theta),
}\\
   {\displaystyle \frac{\partial V}{\partial \psi}=\frac{\gamma\cdot(\sin\varphi\,\cos\psi+\cos\varphi\,\sin\psi\,\cos\theta)}{V},
}\\
   {\displaystyle \frac{\partial k}{\partial \psi}=\frac{2/k-k(\alpha{V}+1) }{(\alpha{V}+1)^2+{z}^2}\cdot{\displaystyle \alpha\frac{\partial V}{\partial \psi} },
}\\ {\displaystyle \frac{\partial z}{\partial \psi}=0.
}
   \end{array}
\end{equation}

Using the derivatives of the kernel obtained above, we can write the first derivatives of Grover's formula of mutual inductance with respect to the appropriate coordinates. Taking into account Eq. (\ref{eq:first rho-der of kernel Grover}),  the $\rho$-derivative of Grover's formula (\ref{eq:GROVER FORMULA}) is
\begin{equation}\label{eq:rho-der GF}
 \frac{\partial\mathrm{M}}{\partial \rho}=\frac{\mu_0}{2\pi}\sqrt{\frac{R_p}{R_s}}\int_{0}^{2\pi}\frac{\partial U}{\partial \gamma} \cdot\Psi(k)+U\cdot\frac{d \Psi(k)}{d k} \cdot\frac{\partial k}{\partial \gamma}\;\;d\varphi.
\end{equation}
Accounting for (\ref{eq:first d-der of kernel Grover}), the $d$-derivative of Grover's formula (\ref{eq:GROVER FORMULA}) can be written as
\begin{equation}\label{eq:d-der GF}
 \frac{\partial\mathrm{M}}{\partial d}=\frac{\mu_0}{2\pi}\sqrt{\frac{R_s}{R_p}}\int_{0}^{2\pi}U\cdot\frac{d \Psi(k)}{d k} \cdot\frac{\partial k}{\partial \Delta}\;\;d\varphi.
\end{equation}
Considering Eq. (\ref{eq:first theta-der of kernel Grover}) and (\ref{eq:first psi-der of kernel Grover}), first derivatives of Grover's formula (\ref{eq:GROVER FORMULA}) of with respect to angular coordinates can be written
\begin{equation}\label{eq:ang coor-der GF}
 \frac{\partial\mathrm{M}}{\partial g}=\frac{\mu_0\sqrt{R_pR_s}}{2\pi}\int_{0}^{2\pi} \frac{\partial U}{\partial g} \cdot\Psi(k)+U\cdot\frac{d \Psi(k)}{d k} \cdot\frac{\partial k}{\partial g}\;\;d\varphi,
\end{equation}
where $g=\theta$ and $\psi$, respectively.

The derived formulas (\ref{eq:rho-der GF}), (\ref{eq:d-der GF}) and (\ref{eq:ang coor-der GF}) can be easily  programmed by using, for instance,  \textit{Matlab} language. The file with \textit{Matlab} code with implemented formulas is available for a reader in   supplementary materials to this article.

\bibliography{References}

\end{document}